# Minimization principle for non degenerate excited states (independent of orthogonality to lower lying known approximants)

Naoum C. Bacalis

*Theoretical and Physical Chemistry Institute, National Hellenic Research Foundation Vasileos Constantinou 48, GR-11635, ATHENS, Greece*

http://arxiv.org/abs/1601.04106

The computation of small concise and comprehensible excited state wave functions is needed because many electronic processes occur in excited states. But since the excited energies are saddle points in the Hilbert space of wave functions, the standard computational methods, based on orthogonality to lower lying approximants, resort to huge and incomprehensible wave functions, otherwise, the truncated wave function is veered away from the exact.
The presented variational principle for excited states, $F_n$, is demonstrated to lead to the correct excited eigenfunction in necessarily small truncated spaces. Using Hylleraas coordinates for He $^1S$ $1s2s$, the standard method based on the theorem of Hylleraas - Unheim, and MacDonald, yields misleading main orbitals 1s1s' and needs a series expansion of 27 terms to be "corrected", whereas minimizing $F_n$ goes directly to the corect main orbitals, $1s2s$, and can be adequately improved by 8 terms. $F_n$ uses crude, rather inaccurate, lower lying approximants and does not need orthogonality to them. This reduces significantly the computation cost. Thus, having a correct 1st excited state $\psi_1$, a ground state approximant can be immediately improved toward an orthogonal to $\psi_1$ function. Also higher lying functions can be found that have the energy of $\psi_1$, but are orthogonal to $\psi_1$. $F_n$ can also recognize a "flipped root" in avoided crossings: The excited state, either "flipped" or not, has the smallest $F_n$. Thus, state average is unnecessary. The method is further applied via conventional configuration interaction up to three lowest singlet states of He.

**The problem and the purpose**

The study of excited states is already imperative especially as it concerns reactions, after activation, of stable species, like $CO_2$ or alkanes. First principles studies can only be utilized in truncated Hilbert spaces. Unfortunately, the standard methods of computing excited states in truncated spaces, although perhaps adequate for the energy and for spectroscopy, may yield incorrect wave functions (perhaps with correct energy), misleading for desired proper excitations. Thus, a method is needed (such as the present demonstrated) to yield excited state truncated wave functions that are not veered away from the exact Hamiltonian eigenfunctions. The ability to extend the variational principle to any excited state (without knowledge of the lower-lying exact eigenfunctions) has long been proven to be an inherent property of the

Hamiltonian. The excited state truncated wave function based on the standard method of the Hylleraas and Undheim / MacDonald (HUM) theorem, is in principle incorrect in a more fundamental manner than just being truncated: Its accuracy must be strictly less than the accuracy of the ground state truncated approximant. On the other hand, an energy minimization orthogonally to all lower *approximants* ["orthogonal optimization" (OO)] must lead to a wave function lying lower than, and veered away from, the exact. A minimization principle for excited electronic states of a non-degenerate Hamiltonian in any given symmetry type is presented, that allows their computation to any desired accuracy, independently of the accuracy of the lower-lying states of the same symmetry, therefore without demanding orthogonality to known lower approximants, and, within a given truncated wave function parameter space, can lead to a more correct than HUM or OO approximation of the excited wave function. A demonstration is presented for the first excited state of He $^1$S ($1s2s$) using variationally optimized, optionally state-specific, orbitals in Hylleraas coordinates, while the standard truncated HUM answer, despite the correct energy, has main orbitals $1s1s'$ instead. It is also demonstrated that the principle can be used to identify a "flipped root" near an avoided crossing (useful to guide MCSCF). Beyond the aforementioned demonstration, some results within conventional configuration interaction based on similarly optimized Laguerre type orbitals are also exhibited and compared to relevant literature.

**Introduction**

The standard quantum computational methods fail to explain various phenomena, like the vanishing sticking probability of $O_2$ on Al(111) surface [1]: Density Functional Theory (DFT) and Quantum-Chemistry computations suggest that $O_2$ should rather be prepared *excited* from triplet to *singlet* in order to be adsorbed on Al(111) [2].

But DFT does not compute excited wave functions, and the standard Quantum-Chemistry methods, via truncated wave functions, can *only* approach the correct excited *energy*, but *not* the exact excited *wave function:* Infinitely many wave functions (*orthogonal* to given lower lying truncated approximants) have *exactly* the excited energy (because, for example, between the exact 1st excited state $\psi_1$ and an arbitrary large normalized expansion $\Phi$, supposed to approximate the 1st excited state, the system of equations $\{\langle\Phi|H|\Phi\rangle \equiv E_1, \langle\psi_1|\Phi\rangle \equiv 0\}$ is generally solved by many sets of expansion coefficients of $\Phi$ [cf. Appendix]), and the standard methods, based on the Hylleraas and Undheim / MacDonald (HUM) theorem [3], i.e. approaching the excited *energy* from *above*, are restricted *not* to approach even the *closest* wave function to the exact excited state, while preserving the above *orthogonality*, because the "closest" *lies below* the exact [4,5 and cf. below]. We have developed a "variational principle" for truncated excited wave functions that approaches the exact excited wave function without any restriction (unlike a HUM approximation) [5]. The *"variational principle"* has preliminarily been applied to atoms [5-7]. It can also be used to identify any "root flipping" appearing in practice in standard methods [8].

*Previous attempts*

Various attempts exist in the literature to introduce *minimization principles for excited states*. The first idea was to minimize $\langle\Phi|(H-E)^2|\Phi\rangle$ [9], whose implementation needed an approximate value of E [10] or other assumptions which did not always guarantee convergence to the correct wave function [11]. Some other early attempts turn the Schrödinger equation $(T+V)\Psi = E\Psi$ into an integral equation using the Green's function $G=(E-T)^{-1}$ and adjust $E$ so as to make the eigenvalue $\mu$ of $\langle\Psi|VGV|\Psi\rangle/\langle\Psi|V|\Psi\rangle = \mu \to 1$. This gives an upper bound of the energy even for discontinuous functions [12,13], but for large systems a development of new basis sets is needed. [14] Besides, no reference is made to whether the wave function, unlike HUM, approaches the correct eigenfunction or not. The generalized Brillouin theorem has also been used in a multiconfiguration method, [15] recursively improving the orbitals by adding improved singly excited orbital corrections, thus reducing a sum of (singly excited) Slater determinants into a single one; it prevents "root flipping" and guarantees upper bound property by orthogonality constraints to lower states, which, as shown here (cf. below), does not guarantee correctness of the excited state *wave function*. Another proposed method, based on the Courant minimax principle, is to minimize the sum of all lower secular roots, including the $n^{th}$, [16] by varying the non linear parameters, but again no reference is made to the correctness of the wave functions. Theophilou [17] has shown that $m$ orthonormal trial functions $\psi_i$ of increasing energy obey $\sum_i E_i \leq \sum_i \langle\psi_i|H|\psi_i\rangle$ where $E_i$ are the $m$ lowest exact (non degenerate) energies. To concentrate on an excited state wave function Gross et al. [18] have further shown that $\sum_i w_i E_i \leq \sum_i w_i \langle\psi_i|H|\psi_i\rangle$ where $w_i$ are weights in *decreasing* order $w_1 \geq ... \geq w_i \geq ... \geq w_m \geq 0$. In another method, tested successfully for a one-dimensional anharmonic oscillator, Ftáčnik et al. proposed to vary the (always positive) variance of the Hamiltonian, $\langle\psi|H^2|\psi\rangle - \langle\psi|H|\psi\rangle^2$, following the steepest descent, but gave no information about the converged wave function [19]. Moreover, an orthogonality constrained energy minimization of Ref. [20-22], as explained and demonstrated here (cf. below), leads to an incorrect wave function lying *below* the exact eigenfunction. Chan et. al [23] minimize the difference between the correct Hamiltonian and a parametrized exactly solvable model, with acceptable results for an anharmonic oscillator. Also for an anharmonic oscillator, it is demonstrated [24] that minimizing the sum of the $n$ lower secular (HUM) roots does not provide a reliable $n^{th}$ wave function and is inferior than minimizing unconstrainedly the $n^{th}$ HUM root. As discussed below, the wave function of the latter is veered away from the exact excited state. Hoffmann et al. [25] minimized an energy state-average along directions of negative curvature. The state average is discussed below. Cancès et al. computed the $1^{st}$ excited state as upper bound of the exact energy by deforming the path to the saddle point pass between $-\varphi_0$ and $\varphi_0$ (the ground state).[26] Dorando et al. used density matrix renormalization group [27 and references therein] for excited states (called "targeted") of quasi one dimensional molecules, by state averaging the density matrix (for stability against root flipping), computed either individually, or as a result of state averaging. Finally, Pastorczak et al. [28] minimized the (truncated) Helmholtz free energy $\sum_i \left(Tw_i \ln w_i + w_i \langle\psi_i|H|\psi_i\rangle\right)$ by adopting (normalized) Boltzmann weights $w_i = e^{-\langle\psi_i|H|\psi_i\rangle/T} / \sum_k e^{-\langle\psi_k|H|\psi_k\rangle/T}$ (at "temperature" parameter T), which reduces to the methods of Ref [17 and 18] as special cases.

*Inadequacy of standard methods*

The computation of excited states in a given truncated wave function parameter space is usually based on HUM theorem [3] by optimizing the desired higher root of the secular equation, (at the expense of the quality of the lower roots). However, as exposed below, the optimized higher root is veered away from the exact excited state, another function of the given truncated space being closest to the exact. HUM theorem guarantees that the more the truncated space is improved, the more the higher root approaches the exact, but often such improvements are impracticable.

Towards a remedy, one could attempt to minimize the excited state energy by searching *orthogonally* to pre-computed *approximants* of the lower states ["orthogonal optimization" (OO)], but, as demonstrated below, OO must lead to *lower* than the exact energy, also veered away from the exact excited state. In explaining accurate experiments, proper approximation of the exact excited *wave function* is needed to compute expectation values of other properties whose "1st derivatives", unlike the energy, do not vanish at it (at the exact) [thus giving error $\sim O(\varepsilon)$ instead of $O(\varepsilon^2)$], and in cases of discrepancies from the experiment it is needed to clarify the role of its error. Due to the aforesaid restrictions, extensively analyzed below, these tasks cannot be achieved by either HUM or OO.

*The purpose*

The purpose of this work is to demonstrate that by varying all parameters, linear and non-linear, the presented *minimization principle for excited states* (non-degenerate) [5] can approach the exact excited wave function more correctly than HUM or OO, within the given *truncated* parameter space, [the larger the space the better the three functions (HUM, OO, and the present) approach the exact] and also that, in case of "root-flipping", the flipped root can be identified [that could allow a safe guidance of standard multi-configuration self-consistent field (MCSCF) calculations].

*The minimization principle for excited states*

The *energy* of excited states does not obey any minimization principle because, as a functional of the Hamiltonian eigenstates $|i\rangle, i = 0,1,2,...$ ($E_0 < E_1 < E_2 < ...$), the energy has saddle points there. [29-33] As Shull and Löwdin [34] have shown, the excited states can be calculated without knowing the lower lying eigenfunctions. A recipe to invent a *minimization principle for excited states*, for a related functional, for a non-degenerate Hamiltonian, in some type of symmetry (by considering inversion of the sign of the "downward parabolas") is:

*Recipe*
(a) to expand the approximation $|\phi_n\rangle$ around $|n\rangle$ (assumed real and normalized),
$$|\phi_n\rangle = \sum_{i \neq n} |i\rangle\langle i|\phi_n\rangle + |n\rangle\sqrt{1 - \sum_{i \neq n}\langle i|\phi_n\rangle^2},$$
(b) to write the energy as $\langle \phi_n | H | \phi_n \rangle = E = E_n - L + U$ where the lower, $L$, and the higher, $U$, terms are
$$L = \sum_{i<n}(E_n - E_i)\langle i|\phi_n\rangle^2, \quad U = \sum_{i>n}(E_i - E_n)\langle i|\phi_n\rangle^2,$$
(c) to introduce a *new* related functional by *inverting the sign of L*, $F_n := +L + E_n + U$ (now all terms are higher than $E_n$),
(d) to substitute the unknown quantities $E_n + U = E + L$ into $F_n$,
$$F_n := E + 2L$$

($F_n$ has minimum at $|\phi_n\rangle = |n\rangle$, $F_n\left[|\phi_n\rangle(=|n\rangle)\right] = E_n$),

(e) to make a continuation (cf. Appendix) from all lower $|i\rangle$, $i < n$ (which are unknown), to predetermined acceptable *approximants* $|\phi_i\rangle$ (known), so that $F_n$ depends on solely known functions, $i < n$ (and on the $n^{\text{th}}$ sought), $F_n\left[|\phi_n\rangle;\{|\phi_i\rangle, i < n\}\right]$ (the continuation turns out to introduce an extra factor: the last denominator in Eq. 3 below),

(f) to show that now $F_n$ still has a critical point at $|\phi_n\rangle = |n\rangle$, and

(g) to determine the conditions on $|\phi_i\rangle$, $i < n$, under which $|\phi_n\rangle = |n\rangle$ is still a local minimum, $F_n\left[|\phi_n\rangle(=|n\rangle)\right] = E_n$.

(Incidentally, note that, since $E[\phi_n] + L - E_n = U > 0$, then $E_n - L < E[\phi_n]$, or

$$E_n - \sum_{i<n}(E_n - E_i)\langle i|\phi_n\rangle^2 < E[\phi_n].$$

That is, since $E_n$ is a saddle point, an approximant $|\phi_n\rangle$ is not necessary to lye above $E_n$, it must only lye above $E_n - L$.)

Clearly, no orthogonality to the lower approximants $|\phi_i\rangle$ is assumed, so, $|\phi_n\rangle$ can be optimized to any desired accuracy [*independently* of the accuracy of $|\phi_i\rangle$ that satisfy the conditions in (g) above; Various such $|\phi_i\rangle$ lead to the *same minimum* of $F_n$]. Therefore, it is not necessary that $|\phi_i\rangle$ have been computed in the same basis (in practice just the main orbitals of a configuration interaction (CI) expansion are enough, which makes this method useful, otherwise, if quite large expansions for $|\phi_i\rangle$ were needed, the method would be practically of no use). Since each $F_n, n = 1, 2, ...$ has minimum at $|\phi_n\rangle = |n\rangle$, the optimal functions minimizing each $F_n$ should result orthogonal to each other.

*HUM inadequacy*

In contrast, as it concerns HUM, in subspaces of truncated basis functions an analysis in Ref. [5] shows that the eigenvector $\Phi_1$ of the "2$^{\text{nd}}$ root" according to the HUM theorem [3] is in principle incorrect in a more fundamental manner than "just truncated": It must have *lower quality* than its orthogonal optimized lowest state "1$^{\text{st}}$ root" $\Phi_0$, in the sense that $\langle 0|\Phi_0\rangle^2$ can approach the maximum value (=1) *at will*, $\langle 0|\Phi_0\rangle^2 \to 1$, whereas $\langle 1|\Phi_1\rangle^2$ is restricted to approach a *lower* value of *no more* than $\left(1 - \langle 1|\Phi_0\rangle^2\right)$, or even less:

$$\langle 1|\Phi_1\rangle^2 < \frac{1 - \langle 1|\Phi_0\rangle^2}{1 + \frac{E_1 - E[\Phi_0]}{-E_1}\frac{\langle 1|\Phi_0\rangle^2}{1 - \langle 1|\Phi_0\rangle^2}} < 1 - \langle 1|\Phi_0\rangle^2, \qquad (1)$$

[or, from another aspect, $\langle 1|\Phi_1\rangle^2 \leq 1 - \langle 1|\Phi_0\rangle^2 - \langle 1|\Phi_2\rangle^2 - \cdots - \langle 1|\Phi_N\rangle^2 < 1 - \langle 1|\Phi_0\rangle^2$, (in fact it is even smaller, as seen in Eq. 1, even if there were no other components)], therefore, $\Phi_1$ (and, similarly, every $\Phi_n$) should be rather computed *independently in a richer basis*, where, then, for the same reason, $\Phi_0$ deteriorates, since

$\langle 0|\Phi_0\rangle^2 \leq 1-\langle 0|\Phi_1\rangle^2 -\langle 0|\Phi_2\rangle^2 -\cdots-\langle 0|\Phi_N\rangle^2 < 1-\langle 0|\Phi_1\rangle^2$. For this reason, the HUM wave function in a (habitual) attempt to minimize directly the 2$^{nd}$ HUM root, which inevitably is orthogonal to a *worse* $\Phi_0$, may tend to the correct energy (from above), but it will be *a priori* much more veered away from the exact $|1\rangle$ [see also Fig. 1 of Ref. [8], briefly presented in the Appendix below].

*OO inadequacy*

Note that $E_n$ would have minimum at $|\phi_n\rangle = |n\rangle$ if $|\phi_n\rangle$ *were orthogonal* to all lower *exact* $|i\rangle$ since the "$L$" term of the energy expansion in (b) above would vanish, but $E_n$ would *not* have minimum at $|\phi_n\rangle = |n\rangle$ if $|\phi_n\rangle$ were orthogonal to any of their *approximants* $|\phi_i\rangle$, i.e. one should not diagonalize (minimize [35]) the *energy* in a subspace *orthogonal* to the lower *approximants* $|\phi_i\rangle$ - in the absence of the exact $|i\rangle$, [8] because (consider for example $\phi_1$:), for some $\phi_0$ the *normalized* function, say $|\phi_1^+\rangle$, which is *orthogonal* to the *approximant* $|\phi_0\rangle$ and is *closest to the exact* $|1\rangle$ (i.e. with the largest projection on $|1\rangle$), belongs to their 2-dimensional (2D) space $\{|\phi_0\rangle, |1\rangle\}$ [any component on higher $|n\rangle$ would diminish its (i.e. the normalized function's) projection on $|1\rangle$]:

$$|\phi_1^+\rangle = \frac{|1\rangle - |\phi_0\rangle\langle\phi_0|1\rangle}{\sqrt{1-\langle 1|\phi_0\rangle^2}} \; ; \quad E[\phi_1^+] = E_1 - \frac{E_1 - E[\phi_0]}{1-\langle 1|\phi_0\rangle^2}\langle 1|\phi_0\rangle^2. \qquad (2)$$

In Eq. 2, if, reasonably, $|\phi_0\rangle$ lies low enough so that $E[\phi_0] < E_1$, then $|\phi_1^+\rangle$'s energy has a value of: $E[\phi_1^+] =$ ($E_1$ diminished (−) by the 2$^{nd}$ term which is a positive quantity) [5], i.e. $|\phi_1^+\rangle$ lies *below* the exact $|1\rangle$ and is not a Hamiltonian eigenfunction. Thus, the *lowest* lying function $|\phi_1^{(LE)}\rangle$, *orthogonal* to the *approximant* $\phi_0$, lies even lower, and, inevitably, is veered away from $|\phi_1^+\rangle$ and, therefore, from $|1\rangle$ [see also Fig. 1 of Ref. [8], and Appendix below].

*A remedy*

The above have been proven in Ref. [5]. $F_n$ is given by

$$F_n[\phi_0,\phi_1,\ldots;\phi_n] \equiv E[\phi_n] + 2\sum_{i<n}\frac{\langle\phi_i|H-E[\phi_n]|\phi_n\rangle^2}{E[\phi_n]-E[\phi_i]}\left[1-\sum_{i<n}\langle\phi_i|\phi_n\rangle^2\right]^{-1}, \qquad (3)$$

where $|\phi_n\rangle = |n\rangle$ is a critical point because at $|\phi_n\rangle = |n\rangle$, the 1$^{st}$ term $E[\phi_n]$ *is* saddle and the 2$^{nd}$ term is zero as an overlap (squared) of the Schrödinger equation "dotted" on a function (i.e. on the weighted sum of the *n* lower approximants $\phi_i$). Further, the conditions for $F_n$ to have a local minimum at $|\phi_n\rangle = |n\rangle$, according the standard theorems of calculus (cf. Sylvester theorem), i.e. that the Hessian determinant and the principal minors along the main diagonal be positive, are:

$$A_n^{k<n} > 0, \; A_n^n > 0, \qquad (4)$$

where

$$A_n^n = 2^{n+1} \prod_{i=0}^{n-1} (E_n - E_i)(E[\phi_n^{\perp\{n\}}] - E_n) \left\{ 1 + 2 \left[ \begin{array}{l} n \sum_{i=0}^{n-1} \langle n|\phi_i\rangle^2 + \\ + \sum_{i=0}^{n-1} \frac{(E[\phi_n^{\perp\{n\}}] - E_n)(E[\phi_i^{\perp\{n\}}] - E_n) - (\langle \phi_i^{\perp\{n\}} | H - E_n | \phi_n^{\perp\{n\}}\rangle)^2}{(E_n - E_i)(E[\phi_i^{\perp\{n\}}] - E_n)} \langle \phi_i^{\perp\{n\}} | \phi_i \rangle^2 \\ -2 \sum_{i=0}^{n-1} \sum_{j=i+1}^{n-1} \frac{((E_n - E_j)\langle j|\phi_i\rangle + (E_n - E_i)\langle i|\phi_j\rangle)^2}{(E_n - E_i)(E_n - E_j)} + O[coefficients]^3 \end{array} \right] \right\}$$

is the Hessian determinant at $|\phi_n\rangle = |n\rangle$, and

$$A_n^{k<n} = 2^{k+1} \prod_{i=0}^{k} (E_n - E_i) \left\{ 1 + 2 \left[ \begin{array}{l} (k+1) \sum_{i=0}^{n-1} \langle n|\phi_i\rangle^2 + \sum_{j=k+1}^{n-1} \sum_{i=0}^{k} \frac{E_n - E_i}{E_n - E_j} \langle i|\phi_j\rangle^2 + \sum_{i=0}^{k} \frac{E[\phi_i^{\perp\{n\}}] - E_n}{E_n - E_i} \langle \phi_i^{\perp\{n\}} | \phi_i \rangle^2 \\ - \sum_{j=k+1}^{n-1} \sum_{i=0}^{k} \frac{E_n - E_j}{E_n - E_i} \langle j|\phi_i\rangle^2 - 2 \sum_{i=0}^{k} \sum_{j=i+1}^{k} \frac{((E_n - E_j)\langle j|\phi_i\rangle + (E_n - E_i)\langle i|\phi_j\rangle)^2}{(E_n - E_i)(E_n - E_j)} \end{array} \right] + O[coeff.]^3 \right\}$$

are its principal minors; $|\phi_i^{\perp\{n\}}\rangle$ is the projection of $|\phi_i\rangle$ on the subspace of higher than-$n$ eigenfunctions. (For clarity in the above expressions, care has been taken to be expressed in terms of positive quantities.)

Note that in these expressions the factor of "1" before the "2[square brackets]" in "{1+2[…]}" normally dominates over the coefficients inside the "2[square brackets]", which are normally small if the lower approximants are accurate enough. Therefore, they do not need be extremely accurate, and as long as they fulfill the conditions (4) without being very accurate, then $|\phi_n\rangle = |n\rangle$ is a local minimum of $F_n$. (The computation of the Hessian, as to the number of its negative eigenvalues, [36] is not needed.)

Depending on how close $\langle 1|\phi_0\rangle \to 0$, the energy $E[\phi_1^{(LE)}]$ can take on any value between $E_0$ and $E[\phi_1^+]$. Also, the "2$^{nd}$ root" $|\phi_1^{"2r"}\rangle$, also orthogonal to some $|\phi_0\rangle$ (i.e. the "1$^{st}$" -*deteriorated* - root), is also veered away from $|\phi_1^+\rangle$ of this $\phi_0$ (and, therefore, from $|1\rangle$) since it lies *higher* [3] than $E_1 > E[\phi_1^+]$ [see Fig.1 of [8] and Appendix]. This means that in the truncated parameter space there are always parameters that lead closer to the exact eigenfunction $|1\rangle$ than the minimized 2$^{nd}$ HUM root $|\phi_1^{"2r"}\rangle$. These can be obtained (say $|\phi_1^{(F)}\rangle$) by minimizing $F_1$ (demonstrated below), and generally $F_n$. The wider the truncated subspace, the more the above three functions $|\phi_1^+\rangle$, $|\phi_1^{(LE)}\rangle$ and $|\phi_1^{"2r"}\rangle$ approach $|1\rangle$, as $\langle 1|\phi_0\rangle \to 0$, [8,15,21,37] along with $|\phi_1^{(F)}\rangle$ [which is independent of all $\phi_0$ s that satisfy inequalities (4)]. As explained below, the limit $\langle 1|\phi_0\rangle \to 0$, i.e. $|\phi_0\rangle$ getting more and more orthogonal to $|1\rangle$, is, in principle, sufficient to find $|1\rangle$ by the above three functions; it is not necessary that $|\phi_0\rangle \to |0\rangle$. (Incidentally, if $|\phi_0\rangle$ happened to belong to the unknown subspace of $\{|0\rangle, |1\rangle\}$, then $|\phi_1^+\rangle$ would be obtained by energy minimization (in full space) orthogonally to $|\phi_0\rangle$, and then $|0\rangle, |1\rangle$ would be obtained as 2x2 eigenfunctions in the $\{|\phi_0\rangle, |\phi_1^+\rangle\}$ subspace. Also, obviously, were $|\phi_0\rangle = |0\rangle$, then $|\phi_1^+\rangle = |1\rangle$. And if $|1\rangle$ were known, then a 2x2 diagonalization in the $\{|\phi_0\rangle, |1\rangle\}$

space would leave $|1\rangle$ unaffected, with higher eigenvalue $E_1$, but would improve $|\phi_0\rangle$ making it orthogonal to $|1\rangle$: $(|\phi_0\rangle - |1\rangle\langle 1|\phi_0\rangle)/\sqrt{1-\langle 1|\phi_0\rangle^2}$ with eigenvalue lower than $E[\phi_0]$: $E[\phi_0] - (E_1 - E[\phi_0])\langle 1|\phi_0\rangle^2 / (1 - \langle 1|\phi_0\rangle^2)$ [35]- as demonstrated below.)

The above prove the *existence* of the local minimum at $|\phi_n\rangle = |n\rangle$ when the lower approximants are accurate enough: In practice, if conditions of Eq. (4) are violated, then $F_n \to -\infty$ because, for some $|\phi_i\rangle$ (the highest), the energy difference $E[\phi_n] - E[\phi_i]$ in Eq. (3) may become negative. This actually happens in MCSCF calculations when variational collapse occurs due to the so called MCSCF "root-flipping" [15,21,25-27,29,31,35-43]: In MCSCF, as mentioned above, improvement of the $n^{th}$ root, by improving its orbitals, deteriorates the lower roots; If *these deteriorated roots* were used in $F_n$ (whose derivation, like MCSCF, also demands flatness and "saddleness" of the energy at $|\phi_n\rangle = |n\rangle$), then root-flipping would be rather unavoidable (in both methods MCSCF and $F_n$). However, contrary to MCSCF (which unavoidably computes the desired higher root orthogonal to the *deteriorated* lower roots), $F_n$ does not need the *deteriorated* lower roots.

*Recognizing "flipped roots"*

It is known that root-flipping may be avoided by an appropriate representation of the excited state, [36] - which may not be known in advance. In $F_n$, by using fixed rather accurate $|\phi_i\rangle$s, root-flipping is avoided: For example, $F_n$ easily and immediately passes the pathological test posed by Rellich [42,44] of the Hermitian matrix $[[-\sin x, \sin y], [\sin y, \sin x]]$ with easily computed eigenvectors $v_0$, $v_1$, if we use a "rather" accurate fixed expansion, in terms of $\{v_0, v_1\}$, i.e.: $\phi_0 = 0.1 v_1 + v_0 \sqrt{1-0.1^2}$ (where 0.1 is "close" to 0, but not too close) and minimize $F_1$ (i.e. $F_n$ for $n = 1$ in Eq. 3) by optimizing $\phi_1 = a v_1 + v_0 \sqrt{1-a^2}$ with respect to $\{a, x, y\}$ around $(a, x, y) \sim (1, 0, 0)$. (The resulting minimum: i.e. at the angles $x = 0$, $y = 0$, is independent of $a$).

Thus, using $F_n$, the problem of root flipping becomes a matter of accuracy. Also, due to the denominators in Eq. (3), the test function $\phi_n$ should, reasonably, not be close to any of the predetermined lower $\phi_i$s; this can always be accomplished by independently increasing the accuracy of $\phi_n$ for fixed $\phi_i$s, since $F_n$ always remains in the higher branch of the avoided crossing without ambiguity in recognizing the functions [42,25]: In principle, with the help of $F_n$, even if the wave functions are computed in the same basis (like in MCSCF), they can be recognized near the crossing: Consider for example the two lowest states. [see also 40] The eigenfunctions, "roots", depend on variational parameters **p** (both linear and non-linear) to be optimized, and say the optimal eigenfunctions are $\Psi_0(\mathbf{p}_0), \Psi_1(\mathbf{p}_1)$. For example, in 1-electron atomic ion S-states $\Psi_0(r) \sim e^{-p_0 Z r}$, $\Psi_1(r) \sim (1 - p_2 Z r / 2) e^{-p_1 Z r / 2}$, $\mathbf{p} = (p_0, p_1, p_2)$, so that at their energy minimum $\mathbf{p}_0 = (1, 0, 0)$, $\mathbf{p}_1 = (0, 1, 1)$. Now, since at $\mathbf{p}_0$: $\Psi_0 \simeq \psi_0$ is the lowest lying, whereas at

$\mathbf{p}_1 : \Psi_1 \simeq \psi_1$ is *excited* - and sought - (the excited is called $\psi_1$), therefore, $\psi_1$ lies *higher than the (deteriorated) lowest lying* ($\Psi_0$) at $\mathbf{p}_1$, both *optimized* roots (close to the exact *eigenfunctions with* $E_0 < E_1$), must obey $E_0 \simeq E[\Psi_0] < E[\Psi_1] \simeq E_1$ (at *both* $\mathbf{p}_0$ and $\mathbf{p}_1$). This means that, at $\mathbf{p}_0$, since $\Psi_0(\mathbf{p}_0) \simeq \psi_0$ is the lowest, we should have that: $E_0 \cong E[\Psi_0(\mathbf{p}_0)] < E[\Psi_1(\mathbf{p}_0)(\text{deteriorated})] \approx E_1$, and at $\mathbf{p}_1$, since $\Psi_1(\mathbf{p}_1) \simeq \psi_1$ is excited, not the lowest, then $E_0 \approx E[\Psi_0(\mathbf{p}_1)(\text{deteriorated})] < E[\Psi_1(\mathbf{p}_1)] \cong E_1 > E_0$. Let us call the whole union of the $\mathbf{p}$-regions where $E[\Psi_0(\mathbf{p})] < E[\Psi_1(\mathbf{p})]$: "*in front of*" the crossing or "*before*" root-flipping. Thus, both $\mathbf{p}_0$ and $\mathbf{p}_1$ are located "*before*" root-flipping. In the variational parameter $\mathbf{p}$-space, "root flipping" $E[\Psi_0(\mathbf{p})] > E[\Psi_1(\mathbf{p})]$ means that the parameters $\mathbf{p}$ are such that the excited state $\Psi_1$, which is sought, has, *behind* the crossing, lower energy than $\Psi_0$ *there*, i.e. the lower "root" consists mainly, i.e. has mainly the characteristics, of the excited state wave function *there*, which we must detect and recognize, in order to extrapolate *it* (i.e. the lower "root" *there*, $\Psi_1$,) by some method, e.g. by quasi-Newton, to regions *in front of* the crossing, close to $\mathbf{p}_1$ (where the unknown optimal $E[\Psi_1(\mathbf{p}_1)]$ must occur - higher than $E[\Psi_0(\mathbf{p}_1)]$).

Near the crossing, let the indices "–"/"+" indicate "just before"/"just behind" the crossing, so that, near the crossing, the higher "2$^{nd}$ root", $\Phi_1^{"2r"}$, is: $\Phi_{1-}^{"2r"} = \Psi_{1-}$, $\Phi_{1+}^{"2r"} = \Psi_{0+}$, and the lower "1$^{st}$ root", $\Phi_0^{"1r"}$, is: $\Phi_{0-}^{"1r"} = \Psi_{0-}$, $\Phi_{0+}^{"1r"} = \Psi_{1+}$, while $\Psi_0$ and $\Psi_1$ are continuous: $\Psi_{0-} = \Psi_{0+} = \Psi_0$ and $\Psi_{1-} = \Psi_{1+} = \Psi_1$. Thus, using a fixed lower approximant $\phi_0$, we have for the higher "2$^{nd}$ root", $F_1[\phi_0; \Phi_{1-}^{"2r"}] = F_1[\phi_0; \Psi_1]$, $F_1[\phi_0; \Phi_{1+}^{"2r"}] = F_1[\phi_0; \Psi_0]$, and similarly for the lower 1$^{st}$ root, $F_1[\phi_0; \Phi_{0-}^{"1r"}] = F_1[\phi_0; \Psi_0]$, $F_1[\phi_0; \Phi_{0+}^{"1r"}] = F_1[\phi_0; \Psi_1]$, which are anyway recognizable from (known) $\mathbf{p}$-points a little farther from the crossing, where unambiguously $F_1[\phi_0; \Phi_1^{"2r"}] = F_1[\phi_0; \Psi_1]$ and $F_1[\phi_0; \Phi_0^{"1r"}] = F_1[\phi_0; \Psi_0]$. Now, the fixed $\phi_0$, independent of the presently varied parameters ($\phi_{0-} = \phi_{0+}$), already optimized at its own energy minimum, is close to $\Psi_0$, i.e. $\langle \Psi_0 | \phi_0 \rangle^2 \sim 1 > \langle \Psi_1 | \phi_0 \rangle^2 \sim 0$, so, near the crossing, the denominators in Eq. 3 are $[1 - \langle \Psi_1 | \phi_0 \rangle^2]^{-1} < [1 - \langle \Psi_0 | \phi_0 \rangle^2]^{-1}$ (normally $\ll$ holds), while the numerators in Eq. 3, $\langle \phi_0 | H | \Psi_0 \rangle - E[\Psi_0]\langle \phi_0 | \Psi_0 \rangle$, $\langle \phi_0 | H | \Psi_1 \rangle - E[\Psi_1]\langle \phi_0 | \Psi_1 \rangle$ remain finite, and since, in optimizing the 2$^{nd}$ "root", the 1$^{st}$ "root" deteriorates -if the (supposedly good) lower approximant $\phi_0$ is better than the deteriorated $\Phi_0^{"1r"}$ there, i.e. if $E[\phi_0] < E[\Phi_0^{"1r"}] \lesssim E[\Phi_1^{"2r"}]$- the 2$^{nd}$ terms (i.e. the sums $2\sum ...$ – the "annexations", so to speak, to the energy) of both $F_1[\phi_0; \Phi_1^{"2r"}]$ and $F_1[\phi_0; \Phi_0^{"1r"}]$ are positive (cf. Eq. 3), that of $\Phi_0$ being normally larger than that of $\Phi_1$, due to the smaller denominator. Therefore, in passing the crossing $F_1[\phi_0; \Phi_1^{"2r"}]$ jumps up from $F_1[\phi_0; \Psi_1]$ to $F_1[\phi_0; \Psi_0]$ (normally high upward):

$$F_1\left[\phi_0;\Phi_{1-}^{"2r"}\right]=F_1\left[\phi_0;\Psi_1\right]\ll F_1\left[\phi_0;\Psi_0\right]=F_1\left[\phi_0;\Phi_{1+}^{"2r"}\right],$$

while $F_1\left[\phi_0;\Phi_0^{"1r"}\right]$ jumps down from $F_1\left[\phi_0;\Psi_0\right]$ to $F_1\left[\phi_0;\Psi_1\right]$ (normally high downward):

$$F_1\left[\phi_0;\Phi_{0-}^{"1r"}\right]=F_1\left[\phi_0;\Psi_0\right]\gg F_1\left[\phi_0;\Psi_1\right]=F_1\left[\phi_0;\Phi_{0+}^{"1r"}\right].$$

In both cases

$$F_1\left[\phi_0;\Psi_1\right]\ll F_1\left[\phi_0;\Psi_0\right].$$

Hence, near the crossing, if $E[\phi_0]<E\left[\Phi_0^{"1r"}\right]\lesssim E\left[\Phi_1^{"2r"}\right]$, then $\Psi_1$ *is recognized to be the "root" that has the lowest $F_1$* (or, anyway, is a continuation of an unambiguous value of $F_1[\phi_0;\Psi_1]$ well before the crossing - but relying in just the "unambiguous continuation", without computing $F_1$, is not always successful). A demonstration of the recognition is presented below for 1-electron atomic ion S-states. (Generally, the recognition of $\Psi_1$ could be used in MCSCF to feed the next iteration. Incidentally, note that "state averaging" *at* the crossing does not provide any information about the correct optimal points **p**$_0$ and **p**$_1$ that must be located "*before*" the crossing.) [cf. Appendix] The above hold as long as the chosen $\phi_0$ satisfies Eq. 4, otherwise it should be improved.

*Minimizing $F_n$*

To minimize $F_n$, the implementation does not need to check in practice any further conditions on $F_n$, implied by Sylvester's theorem, beyond the usual *H*-diagonalization, because the numerator of the 2$^{nd}$ term in Eq. 3 is just Schroedinger's equation for $|\phi_n\rangle$ (projected on each $|\phi_i\rangle$), which vanishes at the minimum $|\phi_n\rangle=|n\rangle$. Thus, having computed $|\phi_n\rangle$ (the lowest above the known and fixed $|\phi_i\rangle$) and, independently, $\langle\phi_i|\phi_n\rangle$, $\langle\phi_i|H|\phi_n\rangle$, then, instead of $E_n$, the value of $F_n$ (immediately calculated by Eq. 3) needs be varied (minimized).

*Immediate improvement of $\phi_0$ via a good approximant of $\psi_1$*

If $\phi_1$, computed in a richer basis, has better quality than a previously determined $\phi_0$, and $\phi_0$ has better quality than (the deterioreted) $\phi_1$'s "1$^{st}$ root" then $\phi_0$ can be immediately improved without any full optimization of the 1$^{st}$ root in the richer basis - by diagonalizing the 2x2 Hamiltonian in their 2D subspace: First, by diagonalizing, we open the gap between $\phi_0$ and $\phi_1$, toward a better function: $\phi_0'$; then, by using any function $\Phi$ *orthogonal to both* $\phi_0'$ and $\phi_1$, we open the gap between $\phi_0'$ and $\Phi$, toward a better function: $\phi_0''$; and so on, until no further improvement occurs. If either of $\phi_1$, or $\phi_0$, were indeed an eigenfunction, then it would be unaffected by the 2x2 diagonalization.

*Criteria of closeness to the exact wave function*

Thus, the above 2x2 diagonalization is a simple **1$^{st}$ criterion** as to how close our functions are to the eigenfunctions. A **2$^{nd}$ criterion** is that in any, whatsoever, calculation of $\phi_n$, the functional $F_n$ can be computed, and it should be a local minimum at $|\phi_n\rangle = |n\rangle$ (if it is not a local minimum, then $|\phi_n\rangle \neq |n\rangle$). A **3$^{rd}$ criterion** is that the energy $E$ (saddle point), just below the minimum of $F_n$, should be a local maximum for at least one of the $\phi_n$'s parameters.

*The demonstration plan*

The above are demonstrated below for He $^1S$ (1$s^2$, and 1$s$2$s$), using Hylleraas variables $s = r_1 + r_2$, $t = r_1 - r_2$ and $u = |\vec{r}_1 - \vec{r}_2|$ [3], by establishing rather accurate basis-functions out of variationally optimized state-specific Laguerre-type orbitals[45], whose polynomial coefficients and exponents are optimized, allowing few term (small-size) series expansions in terms of $s^i (t^2)^j u^k$: E.g. by selecting 24 terms up to $0.001\, s^2 (t^2)^2 u^3$, $E_0 \simeq -2.90372$ a.u. is obtained, but for demonstration reasons all 27 terms will be used up to $s^2 (t^2)^2 u^2$, ($E_0 \simeq -2.90371$ a.u., $E_1 \simeq -2.14584$ a.u.), along with all 8 terms up to $s^1 (t^2)^1 u^1$ whose $\phi_0$ will be immediately improved via the 27-term $\phi_1$. (Pekeris' 9-term $\phi_1$ is still unbound; he reports larger than 95 terms wave functions [46] - indicating that the present optimized Laguerre-type orbitals reduce the size efficiently.) Evidently, if, instead of $F_1$, we minimized $E_1$, the 2$s$ Laguerre-type orbital $(1 - ar) e^{-\zeta_1 r}$ would collapse to 1$s$ ($a \to 0$, $\zeta_1 \to \zeta_0$). This is a transparent example of $E_1$'s "saddleness" at $|\phi_1\rangle = |1\rangle$.

In the next section the formalism is described, including *previously unreported matrix elements of* $\langle\phi_i|\phi_n\rangle$, $\langle\phi_i|H|\phi_n\rangle$ *in Hylleraas coordinates*. Then, the results are given, including the immediate improvement of $\phi_0$ via a more accurate $\phi_1$. Finally, the last section shows a demonstration of recognizing a "flipped root" near the crossing. After the demonstration, a conventional CI computation is presented.

**Formalism**

For two-electron atomic ions of nuclear charge $Z$ the wave function will consist of a single Slater determinant multiplied by a truncated power series of $s = r_1 + r_2$, $t = r_1 - r_2$ and $u = |\vec{r}_1 - \vec{r}_2|$: $\sum_{i_s, i_t, i_u = 0}^{n_s, n_t, n_u} c_{i_s, i_t, i_u} s^{i_s} t^{2 i_t} u^{i_u}$. Due to the spin antisymmetry, the Slater determinant is reduced to a symmetric sum of products, and the power series to a symmetric function of $t$, i.e. of $t^2$. And since the Hamiltonian is also symmetric, the $t$-integrals could be evaluated only for $t > 0$ (eventually multiplied by 2), so that, with volume element $2\pi^2 (s^2 - t^2) u\, dt\, du\, ds$, the limits of integration be $0 < t \leq u \leq s < \infty$ [47].

The spin-orbitals will be composed of Laguerre-type radial orbitals whose polynomial coefficients are treated as variational parameters. [45] For low-lying *singlet* states, only $s$- such orbitals will be considered. Thus, the spatial orbitals will be

$$\chi(n, r; z_n, \{a_{n,k}\}) = \frac{4\sqrt{\pi} \sqrt{(n-1)! n!}}{n^2} z_n^{3/2} \sum_{k=0}^{n-1} \frac{a_{n,k} (-2 r z_n / n)^k e^{-r z_n / n}}{k!(k+1)!(n-k-1)!}$$

where the variational parameters are: $z_n$, and $a_{n,k}$-factors which are expected to have values near 1 (for $k = 0$, $a_{n,0} \equiv 1$) for state-specific functions, allowing also the possibility for non-state-specific description: $a_{n,0} \equiv 1, a_{n,k>0} = 0$. (One would expect the state-specific functions to be more accurate, but in wider parameter space they turn out to be slightly less accurate than the free non-state-specific functions.) The prefactors assert orbital orthonormality for one-electron ions (all $z_n$ = Z and all $a_{n,k}$ =1). With these parameters the mean distance of the electron from the nucleus is separated into a product of a $z_n$ part and an $a_{n,k}$ part, e.g., $\langle r \rangle_s = 3/2z_1$, $\langle r \rangle_{2s} = (3/z_2)(1 - 4a_{2,1} + 5a_{2,1}^2)/(1 - 3a_{2,1} + 3a_{2,1}^2)$, etc. The $a_{n,k}$ part, as a function of $\langle z_n r \rangle$ between 2 and 6 a.u., consists of three branches [cf. Fig.1 (a)]. In the 1$^{st}$ branch $(0 \leq a_{2,1} < 0.33)$ the orbital behaves like 1s with practically negligible negative part [cf. Fig.1 (b)]. In the 2$^{nd}$ branch $(0.33 < a_{2,1} < 1)$ it has one clearly distinct node [cf. Fig.1 (c)]. In the 3$^{rd}$ branch, with a node very close to the nucleus, it is rather unphysical. Similar results hold for the RMS distance $\langle r^2 \rangle^{1/2}$, with almost the same branch separations [cf. Fig.1 (a)]. In fact, around $a_{2,1} \approx 0.33$, where there is an ambiguity about 1s or 2s behavior, some higher Hamiltonian eigenvalues [cf. below], corresponding to a vicinity of meaningless (unphysical) double roots, may be complex, in state-specific description, and the calculation will be abandoned there. Therefore, the relevant $a_{2,1}$ regions will be approximately $(0 \leq a_{2,1} \lesssim 0.27)$ for clear 1s behavior, and $(0.35 \lesssim a_{2,1} \lesssim 2)$ for clear 2s behavior - either as a 2s orbital with a more remote node (small $a_{2,1}$), or as a correlation to the 1s orbital with a node near the 1s maximum (larger $a_{2,1}$). More large $a_{2,1}$ values, where a node is permanently close to the nucleus [cf. Fig. 1(a,d)], will not be tried - as well as values of ambiguous 1s/2s state-specific behavior, around $a_{2,1} \approx 0.33$. Similar remarks apply to higher order orbitals. The power series will be truncated at most up to the 2$^{nd}$ power, $s^2(t^2)^2 u^2$, sufficient for the purposes of the present demonstrations.

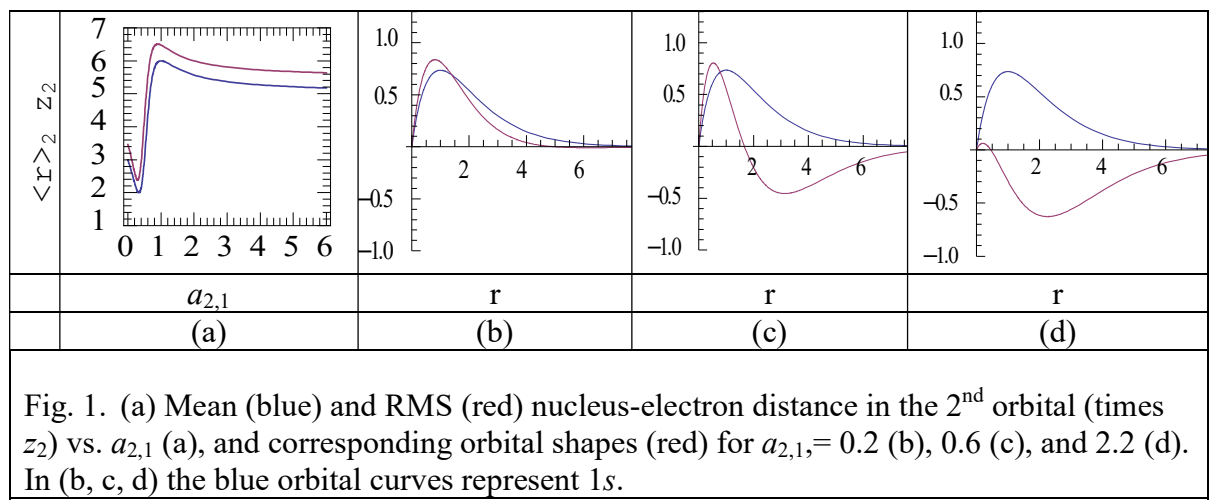

Fig. 1. (a) Mean (blue) and RMS (red) nucleus-electron distance in the 2$^{nd}$ orbital (times $z_2$) vs. $a_{2,1}$ (a), and corresponding orbital shapes (red) for $a_{2,1}$= 0.2 (b), 0.6 (c), and 2.2 (d). In (b, c, d) the blue orbital curves represent 1s.

The two-electron Hamiltonian in terms of the s, t, u variables is given, e.g., in Refs. [47] or [48], but all terms can be reduced to monomial sums by:

(i) selecting $r_1$ or $r_2$ via the sum $(1-q)r_1 + qr_2$, $(q = 0,1)$, (so that the symmetric sum of products be expressed as a monomial sum, e.g. $\sum_{q=0}^{1} \psi((1-q)r_1 + qr_2) \chi(qr_1 + (1-q)r_2)$ ),

(ii) by binomial expanding $r_{1,2}^k \sim (s \pm t)^k$,

(iii) by using $(s^2 - t^2)\varphi = \sum_{p=0}^{1}(-1)^p s^{2(1-p)} t^{2p} \varphi$ and

(iv) by writing the required derivatives as $\frac{d}{dx}(x^n e^{-zx}) = \sum_{b=0}^{1}[n(1-b) - zb]x^{n-1+b}e^{-zx}$.

Thus, the total symmetric two-electron function is written as

$$\phi(\vec{r}_1, \vec{r}_2) = \Phi^{(n_{1\Phi}, n_{2\Phi}, n_{s\Phi}, n_{t\Phi}, n_{u\Phi})}(s,t,u) = \frac{1}{\pi}\sqrt{(n_{1\Phi}-1)!n_{1\Phi}!}\sqrt{(n_{2\Phi}-1)!n_{2\Phi}!}\left(1 - \frac{1}{2}\delta_{n_{1\Phi},n_{2\Phi}}\right)$$

$$\sum_{i_{s\Phi}=0}^{n_{s\Phi}}\sum_{i_{t\Phi}=0}^{n_{t\Phi}}\sum_{i_{u\Phi}=0}^{n_{u\Phi}}\sum_{k_{2\Phi}=0}^{n_{2\Phi}-1}\sum_{k_{1\Phi}=0}^{n_{1\Phi}-1}\sum_{q_\Phi=0}^{1}\sum_{j_{2\Phi}=0}^{k_{2\Phi}}\sum_{j_{1\Phi}=0}^{k_{1\Phi}}\left[(-1)^{j_{1\Phi}+k_{1\Phi}+k_{2\Phi}}(2q_\Phi - 1)^{j_{1\Phi}+j_{2\Phi}} n_{1\Phi}^{-2-k_{1\Phi}} n_{2\Phi}^{-2-k_{2\Phi}} z_{\Phi,n_{1\Phi}}^{\frac{3}{2}+k_{1\Phi}} z_{\Phi,n_{2\Phi}}^{\frac{3}{2}+k_{2\Phi}}\right.$$

$$\left. e^{-s\left(\frac{z_{\Phi,n_{1\Phi}}}{2n_{1\Phi}}+\frac{z_{\Phi,n_{2\Phi}}}{2n_{2\Phi}}\right)-t\left(\frac{1}{2}-q_\Phi\right)\left(\frac{z_{\Phi,n_{1\Phi}}}{n_{1\Phi}}-\frac{z_{\Phi,n_{2\Phi}}}{n_{2\Phi}}\right)} s^{i_{s\Phi}-j_{1\Phi}-j_{2\Phi}+k_{1\Phi}+k_{2\Phi}} t^{2i_{t\Phi}+j_{1\Phi}+j_{2\Phi}} u^{i_{u\Phi}} a_{\Phi,n_{1\Phi},k_{1\Phi}} a_{\Phi,n_{2\Phi},k_{2\Phi}} c_{\Phi,i_{s\Phi},2i_{t\Phi},i_{u\Phi}}\right] /$$

$$\left[j_{1\Phi}! j_{2\Phi}!(k_{1\Phi}+1)!(k_{1\Phi}-j_{1\Phi})!(k_{2\Phi}+1)!(k_{2\Phi}-j_{2\Phi})!(n_{1\Phi}-1-k_{1\Phi})!(n_{2\Phi}-1-k_{2\Phi})!\right]$$

where $n_{i_\Phi}$ is the degree of the associated Laguerre polynomial of the $i^{th}$ spin-orbital in the $\Phi$-Slater determinant (which is multiplied by the power series expansion), so that the total wave function is characterized by the set of integers $(n_1, n_2, n_s, n_t, n_u)$ (where $n_i - 1$ is the number of nodes of the $i^{th}$ polynomial). Thus, the ground state's, $1s^2$, approximant is characterized: by $(1,1,n_s,n_t,n_u)$, the $1^{st}$ excited state's, $1s2s$, approximant: by $(1,2,n_s,n_t,n_u)$, or equivalently by $(n_1 = 2, n_2 = 1, n_s, n_t, n_u)$, in state-specific description, or also by the $2^{nd}$ root of "$(1,1,n_s,n_t,n_u)a = 0$" in non state-specific description (where "$a = 0$" means that all parameters $a_{n,k\neq 0} = 0$), and these can be calculated either by using (minimizing) the $F$ functional - indicated by an index F, or not.

By applying Green's theorem in the integrals, in order to deal with first derivatives, $\Phi_s, \Phi_t, \Phi_u$, instead of Laplacians, the Hamiltonian and overlap matrix elements are:

$$\langle \Phi | H | \Psi \rangle = 2\pi^2 \int_0^\infty \int_0^s \int_0^u [-4Z\, s\, u\, \Phi\, \Psi + \sum_{p=0}^{1}(-1)^p (s^{2(1-p)} t^{2p} u (\Phi_s \Psi_s + \Phi_t \Psi_t + \Phi_u \Psi_u) +$$

$$u^{2(1-p)} t^{2p} s (\Phi_u \Psi_s + \Phi_s \Psi_u) + s^{2(1-p)} u^{2p} t (\Phi_u \Psi_t + \Phi_t \Psi_u) + s^{2(1-p)} t^{2p} \Phi \Psi)] dt\, du\, ds$$

and

$$\langle \Phi | \Psi \rangle = 2\pi^2 \int_0^\infty \int_0^s \int_0^u \sum_{p=0}^{1}(-1)^p s^{2(1-p)} t^{2p} u\, \Phi\, \Psi\, dt\, du\, ds.$$

The following integrals of the form $J(z_s, z_t; n_s, n_t, n_u) = \int_0^\infty \int_0^s \int_0^u e^{-z_s s - z_t t} s^{n_s} t^{n_t} u^{n_u} dt\, du\, ds$ are needed (previously unreported):

$$J(z_s, 0; 0, 0, 0) = z_s^{-3}$$

$$J(z_s, 0; n_s, 0, 0) = 2^{-1} z_s^{-(n_s+3)} (n_s + 2)!$$

$$J(z_s, 0; 0, n_t, 0) = z_s^{-(n_t+3)} n_t!$$

$$J(z_s, 0; n_s, n_t, 0) = \left(n_t^2 + 3n_t + 2\right)^{-1} z_s^{-(n_s+n_t+3)} (n_s + n_t + 2)!$$

$$J(z_s, 0; 0, 0, n_u) = z_s^{-(n_u+3)} (n_u + 1)!$$

$$J(z_s, 0; n_s, 0, n_u) = (n_u + 2)^{-1} z_s^{-(n_s+n_u+3)} (n_s + n_u + 2)!$$

$$J(z_s, 0; 0, n_t, n_u) = (n_t + 1)^{-1} z_s^{-(n_t+n_u+3)} (n_t + n_u + 1)!$$

$$J(z_s, 0; n_s, n_t, n_u) = (n_t + 1)^{-1} (n_t + n_u + 2)^{-1} z_s^{-(n_s+n_t+n_u+3)} (n_s + n_t + n_u + 2)!$$

$$J(z_s, z_t; 0, 0, 0) = z_s^{-2} (z_s + z_t)^{-1}$$

$$J(z_s, z_t; n_s, 0, 0) = z_t^{-2} \left[ (z_s + z_t)^{-(n_s+1)} + z_s^{-(n_s+2)} \left( (n_s + 1) z_t - n_s z_s \right) \right] n_s!$$

$$J(z_s, z_t; 0, n_t, 0) = z_s^{-2} (z_s + z_t)^{-(n_t+1)} n_t!$$

$$J(z_s, z_t; n_s, n_t, 0) =$$

$$\frac{(n_t+1)!}{z_t^{n_t+2}} \left( \sum_{k=0}^{n_t+1} \frac{z_t^k (n_s+k)!}{k! (z_s + z_t)^{n_s+1+k}} - \frac{n_s!}{z_s^{n_s+1}} \right) - \frac{n_t!}{z_t^{n_t+1}} \left( \sum_{k=0}^{n_t} \frac{z_t^k (n_s+1+k)!}{k! (z_s + z_t)^{n_s+2+k}} - \frac{(n_s+1)!}{z_s^{n_s+2}} \right)$$

$$J(z_s, z_t; 0, 0, n_u) = \left[ z_s^{-(n_u+1)} - (z_s + z_t)^{-(n_u+1)} \right] z_s^{-1} z_t^{-1} n_u!$$

$$J(z_s, z_t; n_s, 0, n_u) = \frac{(n_s + n_u + 1)!}{z_s^{n_s+n_u+2} z_t (n_u + 1)} + \frac{n_u!}{z_t^{n_u+2}} \left( \sum_{k=0}^{n_u} \frac{z_t^k}{k!} \frac{(n_s+k)!}{(z_s+z_t)^{n_s+1+k}} - \frac{n_s!}{z_s^{n_s+1}} \right)$$

$$J(z_s, z_t; 0, n_t, n_u) = \frac{n_t!}{z_s z_t^{n_t+1}} \left( \frac{n_u!}{z_s^{n_u+1}} - \sum_{k=0}^{n_t} \frac{z_t^k (n_u+k)!}{k! (z_s+z_t)^{n_u+1+k}} \right)$$

$$J(z_s, z_t; n_s, n_t, n_u) =$$

$$\frac{z_t^{n_u+1} n_t! (n_s+n_u+1)! - z_s^{n_u+1} n_s! (n_t+n_u+1)!}{(n_u+1) z_t^{n_t+n_u+2} z_s^{n_s+n_u+2}} + \frac{n_t!}{z_t^{n_t+n_u+2}} \sum_{j=0}^{n_t} \frac{(n_u+j)!}{j!} \sum_{k=0}^{n_u+j} \frac{(n_s+k)! z_t^k}{k! (z_s+z_t)^{n_s+1+k}} \ .$$

The $k$- and $j$-sums stem from incomplete $\Gamma$-functions, [49] which, alternatively, can be directly computed [50]. For small $z_t$, due to large alternating terms, direct expansion is more appropriate:

$$J(z_s, (|z_t| \ll 1); n_s, n_t, 0) \simeq \sum_{i=0}^{n_t+5} \frac{(-1)^i}{i!} z_t^i \frac{z_s^{-3-i-n_s-n_t} (2+i+n_s+n_t)!}{(1+i+n_t)(2+i+n_t)}$$

$$J(z_s, (|z_t| \ll 1); n_s, 0, n_u) \simeq \sum_{i=0}^{n_u+5} \frac{(-1)^i}{i!} z_t^i \frac{z_s^{-3-i-n_s-n_u} (2+i+n_s+n_u)!}{(1+i)(2+i+n_u)}$$

$$J(z_s, (|z_t| \ll 1); 0, n_t, n_u) \simeq \sum_{i=0}^{n_t+5} \frac{(-1)^i}{i!} z_t^i \frac{z_s^{-3-i-n_t-n_u} (1+i+n_t+n_u)!}{1+i+n_t}$$

$$J(z_s, (|z_t| \ll 1); n_s, n_t, n_u) \simeq \sum_{i=0}^{n_t+n_u+5} \frac{(-1)^i}{i!} z_t^i \frac{z_s^{-3-i-n_s-n_t-n_u} (2+i+n_s+n_t+n_u)!}{(1+i+n_t)(2+i+n_t+n_u)} \ .$$

*$F_n$ minimization procedure*

Now, assuming that the lower approximants $\{\phi_i, i < n\}$ (needed to compute $F_n$, which does *not* require them to be so accurate) have been computed (independently of each other, and not necessarily orthogonal to each other – orthogonality is *not* required to compute $F_n$), for each (varied) value of the parameters $z_n$ and $a_{n,k}$ the critical points of $F_n$ (saddle points of $E_n$) must be found, in a wider subspace of a richer basis, and $F_n$ be minimized (*without* demanding orthogonality to the lower approximants $\{\phi_i, i < n\}$ : orthogonality among the functions that minimize each $F_i, i < n$, should be an outcome). In finding the saddle points of $E_n$, for certain trial values of the parameters $z_n$ and $a_{n,k}$ the linear part, $\partial \langle \phi_n | H | \phi_n \rangle / \partial c_{i_s, 2i_t, i_u} = 0$, can be solved either by direct variation of the $c$-coefficients, or by reducing to a generalized eigenvalue problem (requiring $\langle \phi_n | \phi_n \rangle$ for normalization). Thus, the *lowest* eigenvalue (1$^{st}$ root) above the known (highest) $E[\phi_{n-1}]$ –and the $c$-coefficients of its eigenvector are substituted in Eq. 3 of $F_n$, and improved values of $z_n$ and $a_{n,k}$ are sought toward a lower value of $F_n$. This process is repeated until minimization of $F_n$.

*Establishing "exact" wave functions $\psi_0$, $\psi_1$ and truncated approximants $\phi_0$, $\phi_1$*

First we establish a quite reliable wave function basis $|\psi_0\rangle \cong |0\rangle$, $|\psi_1\rangle \cong |1\rangle$ by taking 27 terms, i.e. up to $s^2(t^2)^2 u^2$, $(n_s, n_t, n_u) = (2,2,2)$, adequate to achieve coincidence of the HUM and $F_1$ minima: ($E[\psi_0] = -2.90371$ a.u., $z_0 = 1.9549$), ($E[\psi_1^{HUM}] = -2.14584$, $z_{1,1} = 1.8348$, $z_{1,2} = 1.9745$, $a_{1,2,1} = 0$; or $E[\psi_1^{F_1}] = -2.14577$, $z_{1,1} = 1.930501$, $z_{1,2} = 1.827298$, $a_{1,2,1} = 0.799760$; $\langle \psi_1^{F_1} | \psi_1^{HUM} \rangle = 0.99996$) - thus, going up to $(n_s, n_t, n_u) = (2,2,2)$ is rather sufficient - and then we truncate up to $s^1(t^2)^1 u^1$, $(n_s, n_t, n_u) = (1,1,1)$, i.e. 8 terms, in order to exhibit the aforesaid demonstrations.

In $F_1$, a fixed 1-term normalized $\phi_0$ i.e. up to $s^0(t^2)^0 u^0$, $(n_s, n_t, n_u) = (0,0,0)$ ($z_0 = 1.6875$, $E[\phi_0] = -2.84766$ a.u.) is used.

**Results**

For He ($Z = 2$) $^1S$, the exact eigenvalues are [46] $E[0] = -2.90372$ a.u., $E[1] = -2.14597$ a.u., $E[2] = -2.06127$ a.u..

**I.** $(n_s, n_t, n_u) = (2,2,2)$: **27 Terms**

Now first establish a reliable basis $|\psi_0\rangle \cong |0\rangle$, $|\psi_1\rangle \cong |1\rangle$, (27 terms) in order to compare (project on it) the truncated $(n_s, n_t, n_u) = (1,1,1)$ 8-term approximants.

The optimized ground state approximant $0^{(1,1,2,2,2)}$ [cf. monosyllabic "tSzaR" = ψ0000 in TABLE 1] has lowest root energy $E[0^{(1,1,2,2,2)}] = -2.90371$ a.u. and parameters $z_{0,1} = 1.954881$, $\{c_{0,0,0,0}=1, c_{0,0,0,1} = 0.397612, c_{0,0,1,0} = 0.219317, c_{0,0,1,1} = -0.179310, c_{0,1,0,0} = 0.031004, c_{0,1,0,1} = 0.070486, c_{0,1,1,0} = 0.053015, c_{0,1,1,1} = 0.008797, c_{0,0,0,2} = -0.099934, c_{0,0,1,2} = 0.072875, c_{0,1,0,2} = 0.017865, c_{0,1,1,2} = -0.022953, c_{0,0,2,0} = -0.000292, c_{0,0,2,1} = -0.001140, c_{0,1,2,0} = 0.001706, c_{0,1,2,1} = 0.002110, c_{0,2,0,0} = 0.033369, c_{0,2,0,1} = -0.001512, c_{0,2,1,0} = -0.004711, c_{0,2,1,1} = 0.006209, c_{0,0,2,2} = -0.002102, c_{0,1,2,2} = 0.000535, c_{0,2,0,2} = -0.000467, c_{0,2,1,2} = 0.001339, c_{0,2,2,0} = -0.00088, c_{0,2,2,1} = -0.000190, c_{0,2,2,2} = -0.000033\}$, with normalization factor $N = 1/1.74187$. (The optimized 2$^{nd}$ root with the same $z_{0,1}$ is rather high: (tSzaR = ψ0001), $E = -2.1391$ a.u..)

TABLE 1. Energies and overlaps of the computed states. Notation "tSzaR": t = {ψ: 27 terms, Φ: 8 terms}, S = {0: ground state treatment, 1: excited state treatment}, z = {0: $z_1$ only, 1: $z_1,z_2$}, a = {0: a=0, 1: a≠0}, R = {0: 1$^{st}$ root, 1: 2$^{nd}$ root, H: HUM, F: $F_1$}

| Φ = tSzaR | E (a.u.) | (ψ$_0$|Φ) | (ψ$_1$|Φ) | (Φ020H |Φ) |
|---|---|---|---|---|
| ψ0000→**ψ$_0$** | **-2.90371** | | | |
| (ψ0001) | (-2.1391) | | | |
| ψ0100 | -2.90371 | 1. | | |
| (ψ0101) | (-2.1306) | | | |
| ψ110H→**ψ$_1$** | **-2.14584** | 0.000172 | | |
| (ψ1100) | (-2.90327) | | | |
| ψ111F | -2.14577 | 0.000457 | 0.99996 | |
| | | | | |
| φ$_0$ (1-term) | -2.84766 | 0.993 | | |
| | | | | |
| Φ0000 | -2.903121 | 0.999958 | 1.3 10$^{-5}$ | |
| (Φ0001) | (-2.01016) | | | |
| Φ100H | -2.07215 | 1.97 10$^{-5}$ | 0.875 | |
| (Φ1000) | (-2.89748) | | | |
| Φ100F | -2.07215 | | | |
| **Φ020H** | **-2.903123** | | | |
| (Φ0201) | (-2.0196) | | | |
| | | | | |
| Φ120H | -2.14449 | | | 0.0026 |
| (Φ1200) | (-2.8886) | | | |
| Φ120F | -2.14449 | | | |
| Φ121H→**1s1s′** | **-2.14449** | **0.0033** | **0.9986** | |
| Φ121F→**1s2s** | **-2.145152** | **0.00490** | **0.999807** | 0.0186 |

By including the $z_{1,2}$ parameter, but $a_{1,2,1} = 0$, the optimized 1$^{st}$ root $0^{(1,2,2,2,2)a=0}$ (tSzaR = ψ0100) has $E[0^{(1,2,2,2,2)a=0}] = -2.90371$ a.u., $z_{1,1} = 1.93945$, $z_{1,2} = 3.8926$ and corresponding c-expansion coefficients (in the same order as above) $\{c\} = \{1, 0.396427, 0.218472, -0.175615, 0.020277, 0.067631, 0.049624, 0.010133, -0.100055, 0.069971, 0.018931, -0.022001, -0.000431, -0.000782, 0.001552, 0.001955, 0.031631, -0.002881, -0.004778, 0.005708, -0.002076, 0.000528,$

−0.000586, 0.001288, −0.000813, −0.000182, −0.000032}, $N$ = 1/ 3.42055. (The optimized 2$^{nd}$ root with the same $z_{1,1}$, $z_{1,2}$ is rather high again: (tSzaR = ψ0101) $E$ = −2.1306 a.u..)

Since the above two approximants represent the same unique ground state [as seen in Fig. 2, the minimum is very flat, but indeed, within the accuracy of the computation (~ 6 digits[*]), the 2$^{nd}$ function has, as expected, $z_{1,2}$ = 2 $z_{1,1}$, i.e. both represent the same 1$s$ orbital, with {$c$} coefficients very similar to the 1$^{st}$, and $\langle 0^{(1,1,2,2,2)} | 0^{(1,2,2,2,2)a=0} \rangle = 1.$ = (ψ0000|ψ0100)], we shall use the simplest, $0^{(1,1,2,2,2)}$, (tSzaR = ψ0000), as a quite reliable approximant of the ground state wave function ψ$_0$. [Incidentally, observe that because the minimum is very flat, any traditional convergence criterion (e.g. δE < ε = 10$^{-6}$) could be easily satisfied anywhere between 1.5 < $z_{1,1}$, < 2.3, so that any other property that does not have extremum at ψ$_0$ but rather passes from it with a finite slope, would be computed arbitrarily incorrectly. The same comment holds also for the 1$^{st}$ excited state - [cf. below and Fig.2] - which is also quite flat in varying the $z_{1,1}$, $z_{1,2}$ parameters.]

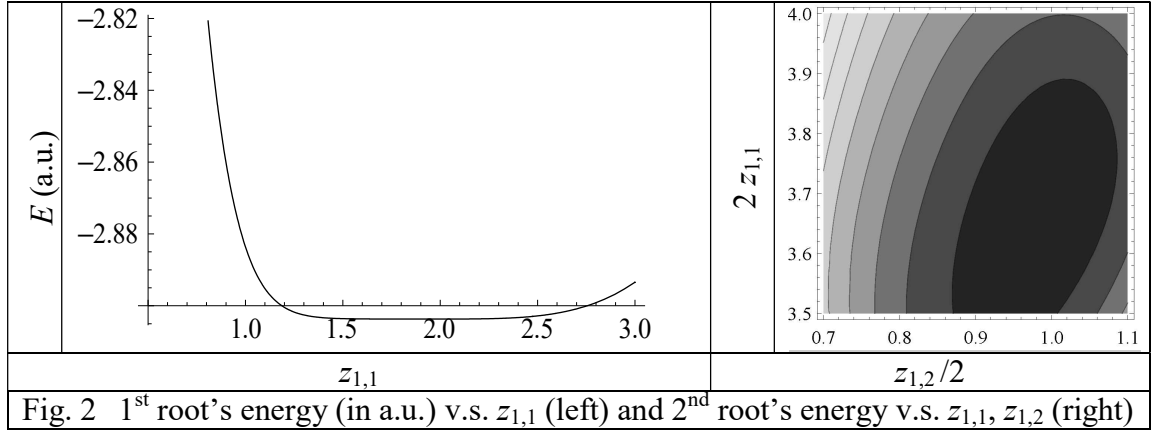

Fig. 2   1$^{st}$ root's energy (in a.u.) v.s. $z_{1,1}$ (left) and 2$^{nd}$ root's energy v.s. $z_{1,1}$, $z_{1,2}$ (right)

The optimized 2$^{nd}$ root $1^{(1,2,2,2,2)a=0}$ has (tSzaR = ψ110H) $E[1^{(1,2,2,2,2)a=0}]$ = −2.14584 a.u., $z_{1,1}$ = 1.8348, $z_{1,2}$ = 1.9745 and corresponding $c$-expansion coefficients (always in the same order as above) {$c$} = {1, 0.332324, −0.001830, −0.071914, −0.428532, −0.085846, −0.055044, 0.014824, −0.069869, 0.008250, 0.030197, −0.001585, −0.006722, 0.001822, 0.001660, −0.000404, 0.020320, −0.009953, −0.010853, −0.001296, −0.000407, 0.000083, −0.002444, 0.000189, −0.000053, −0.000019, −3.252820 x 10$^{-6}$}, $N$ = 1/ 3.78752. (Of course, this is orthogonal to its deteriorated 1$^{st}$ root Φ$_0$$^{1r}$, (tSzaR = ψ1100) $E$ = −2.90327 compared to $E[0^{(1,1,2,2,2)}]$ = −2.90371).

It has overlap $\langle 0^{(1,1,2,2,2)} | 1^{(1,2,2,2,2)a=0} \rangle = 0.000172$ = (ψ0000|ψ110H), while its main two orbitals are as in Fig. 1b, resembling 1$s$,1$s'$, rather than 1$s$,2$s$ [see Fig. 6 below]. However, since it is an eigenfunction of the $c$-coefficient secular equation, with optimized $z$-values, its F-value is the same ($F-E$ = 3 x 10$^{-9}$), indicating that the whole series of this function is rather acceptable.

By allowing $a_{1,2,1} \neq 0$, the optimized 2$^{nd}$ HUM root has $a_{1,2,1} \approx 0$, which introduces a node very far from the nucleus (rendering it *literally* "2"$s$!, but *essentially* 1$s'$) with negligibly better energy than the above with $a_{1,2,1}$ = 0, but also has another minimum, except $a_{1,2,1} \approx 0$, with $E[1^{(1,2,2,2,2)}]$ = −2.126 a.u.. Although $z_{1,1}$ =

---
[*] The computation is performed in 15 digit "double precision" but the co-existence in the secular matrix of very large with small numbers reduces the accuracy to ~ 6 digits

2.106, $z_{1,2}$ = 2.478, $a_{1,2,1}$ = 0.539, which mean that the *main* (without the truncated series expansion) orbital 2s is orthogonal to the *main* 1s, this high lying function is rejected, suggesting that using a (perhaps habitual) criterion *of orthogonality of the main orbitals,* in computing excited states, **is not safe**.

The optimized "$F_1$" root $\Phi_1^F$, (the 1$^{st}$ above the fixed $E[\phi_0]$ = −2.84766 a.u.) has (tSzaR = ψ111F) $E[1^{(1,2,2,2,2)F}]$ = −2.14577 a.u., ($F - E$ = 3 x 10$^{-8}$) a.u., $z_{1,1}$ = 1.930501, $z_{1,2}$ = 1.827298, $a_{1,2,1}$ = 0.799760 and corresponding *c*-expansion coefficients $\{c\}$ = {1., 0.342507, 0.102594, −0.038561, −0.132479, 0.052163, 0.025116, −0.001629, −0.081932, 0.006461, 0.011531, −0.000914, −0.004275, 0.001844, −0.000335, −0.000045, 0.028384, −0.009904, 0.000546, 0.000326, −0.000174, 0.000016, 0.000167, 0.000023, 0.000015, −1.907833 x 10$^{-6}$, −3.366412 x 10$^{-7}$}, $N$ = 1/3.48241. $\Phi_1^F$ is also an eigenfunction of the *c*-coefficient secular equation, with optimized *z*- and *a*-values and its main orbitals are as in Fig. 1c, resembling 1s,2s (cf. Fig 6). It has overlap $\langle 0^{(1,1,2,2,2)} | 1^{(1,2,2,2,2)F} \rangle$ = 0.000457 = (ψ0000|ψ111F) and $\langle 1^{(1,2,2,2,2)a=0} | 1^{(1,2,2,2,2)F} \rangle$ = 0.99996 = (ψ110H|ψ111F). Although their main orbitals differ, their whole series expansions, up to $(n_s, n_t, n_u) = (2,2,2)$, essentially coincide.

*Checking the satisfaction of Schrödinger's equation*

If these two functions $\Phi_1^{2r}$ and $\Phi_1^F$ are close to the exact, they should satisfy Schrödinger's equation for any normalized "$\phi_0$" not orthogonal to $\Phi_1$:

$$Schr \equiv \langle \phi_0 | (H | \Phi_1 \rangle - E[\Phi_1] | \Phi_1 \rangle) = \langle \phi_0 | H | \Phi_1 \rangle - E[\Phi_1] \langle \phi_0 | \Phi_1 \rangle \sim 0.$$

(In principle it might be possible for some $\langle \phi_0 |$ to be accidentally almost orthogonal to *both* $|\Phi_1\rangle$ *and* $H|\Phi_1\rangle$, making *Schr* artificially small without $\Phi_1$ being close to the exact, with a danger to pull $F_1$ below $E_1$: $E < F_1 < E_1$. Such an accidental (improbable) case can be easily checked via another $\phi_0$ for the same $\Phi_1$.)

Using the above fixed $\phi_0$, the *Schr* difference equals ~3 10$^{-5}$ for $\Phi_1^{2r}$ and ~8 10$^{-5}$ for $\Phi_1^F$. Fig. 3 shows the *Schr* difference for both $\Phi_1^{2r}$ and $\Phi_1^F$, along with their

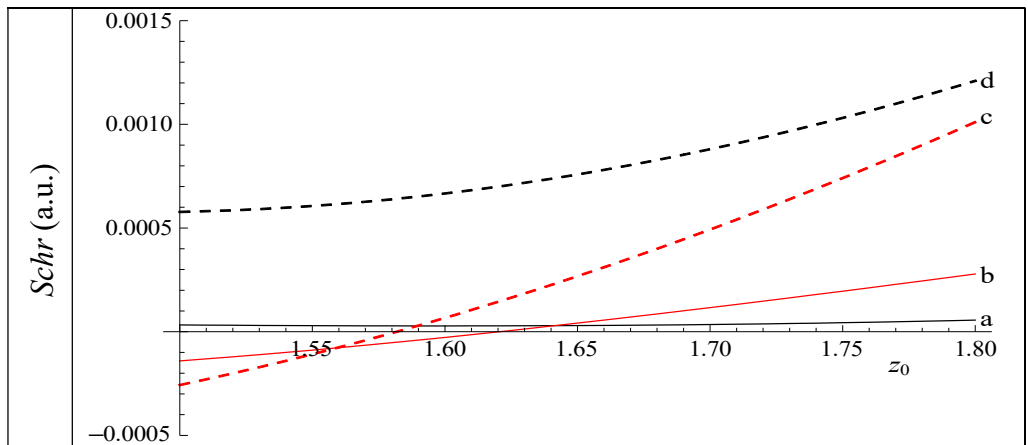

Fig. 3 The difference $Schr \equiv \langle 0^{(1,1,0,0,0)} | (H - E[\Phi]) | \Phi \rangle$ vs. the exponent $z_0$ of normalized $0^{(1,1,0,0,0)}$, for optimized $\Phi$ = (a): $1^{(1,2,2,2,2)a=0}$ HUM 2$^{nd}$ root, (b): $1^{(1,2,2,2,2)F,a}$, and (cf. below) (c): $1^{(1,2,1,1,1)F,a}$, (d): $1^{(1,2,1,1,1)a=0}$ HUM 2$^{nd}$ root.

more truncated versions (cf. below) $1^{(1,2,1,1,1)a=0}$ (*Schr* ~8 10$^{-4}$) and $1^{(1,2,1,1,1)F}$ (*Schr* ~4

$10^{-4}$), for various values of the exponent $z_0$ of (normalized) $\phi_0$. Indeed, contrary to the more truncated versions, $\Phi_1^{2r}$ has slightly better *Schr* values than $\Phi_1^F$.

Thus, we have established a quite reliable basis (27 terms), adequate for the attempted demonstration, by adopting $\{\psi_0 = 0^{(1,1,2,2,2)}, \psi_1 = 1^{(1,2,2,2,2)a=0}\}$, since the state-specific expansion, despite its more reasonable main orbitals, is slightly inferior in *Schr* than the expansion of the HUM 2$^{nd}$ root.

## II. $(n_s, n_t, n_u) = (1,1,1)$: 8 Terms

The optimized ground state 8-term approximant $0^{(1,1,1,1,1)}$ has lowest root energy (cf. TABLE 1: tSzaR = Φ0000) $E[0^{(1,1,1,1,1)}] = -2.903121$ a.u. and parameters $z_{0,1} = 1.84250$, $c_{0,0,0,0} = 1$, $c_{0,0,0,1} = 0.290798$, $c_{0,0,1,0} = 0.190212$, $c_{0,0,1,1} = -0.0765151$, $c_{0,1,0,0} = 0.00688611$, $c_{0,1,0,1} = 0.0139948$, $c_{0,1,1,0} = 0.016833$, $c_{0,1,1,1} = 0.0120625$, normalization constant $N = 1/1.47861$, whereas its orthogonal 2$^{nd}$ root $1^{\left[0^{(1,1,1,1,1)}\right]}$ (same $z_{0,1}$), lies too high: (tSzaR = Φ0001) $E[1^{\left[0^{(1,1,1,1,1)}\right]}] = -2.01016$ a.u. This ground state truncation, $0^{(1,1,1,1,1)}$, having $\langle\psi_0|0^{(1,1,1,1,1)}\rangle = 0.999958$ and $\langle\psi_1|0^{(1,1,1,1,1)}\rangle = 1.3 \cdot 10^{-5}$, is nearly "perfect". However, for the 1$^{st}$ excited state, as seen below, this truncation (1,1,1,1,1) is not adequate: A richer function (1,2,1,1,1) is needed.

The optimized 8-term 2$^{nd}$ HUM root $1^{(1,1,1,1,1)}$ (Fig. 3), has (tSzaR = Φ100H) $E[1^{(1,1,1,1,1)}] = -2.07215$ a.u., $z_{1,1} = 1.44234$ and corresponding $c$-expansion coefficients $\{c\} = \{1, 0.199163, 0.0540545, 0.0273041, -0.418026, -0.0402484, -0.0720681, -0.040584\}$, $N = 1/1.3566$, while its orthogonal 1$^{st}$ root $0^{\left[1^{(1,1,1,1,1)}\right]}$ (same $z_{1,1}$), is, of course, deteriorated: (tSzaR = Φ1000) $E[0^{\left[1^{(1,1,1,1,1)}\right]}] = -2.89748$ a.u., $\{c\} =$

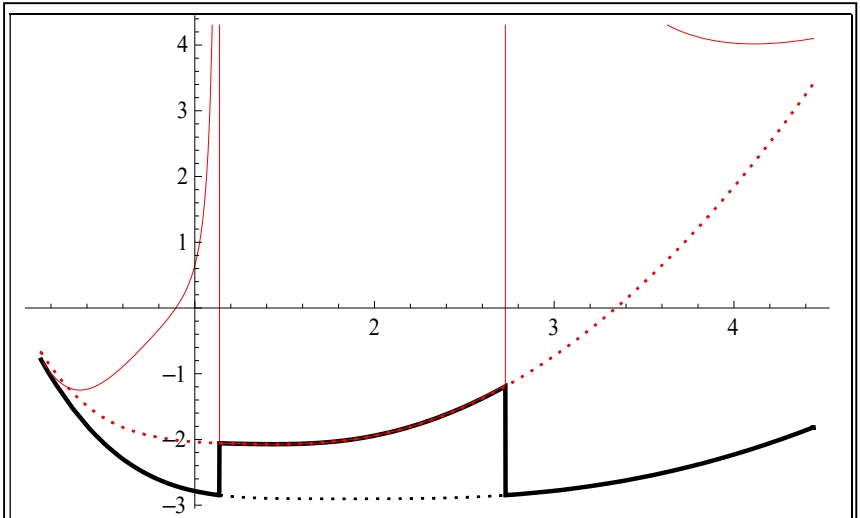

Fig. 4 Energy $E[0^{(1,1,1,1,1)}]$ of the ground and $E[1^{(1,1,1,1,1)}]$ of the 1$^{st}$ excited state approximants (dotted) along with $F_1$ (red) and the first root above the energy of the fixed $\phi_0$ (black). The local (here global) minimum of $F_1$ is near the 2$^{nd}$ HUM root (in lack of other parameters).

$\{1, 0.145873, 0.160241, -0.0416745, -0.245673, -0.00154526, -0.0197971, 0.00659241\}$. Despite the high energy, the overlap between the normalized

independently optimized HUM approximants is $\langle 0^{(1,1,1,1,1)} | 1^{(1,1,1,1,1)} \rangle = -1.97349 \times 10^{-5} = (\psi_0 | \Phi 100H)$, which means that $1^{(1,1,1,1,1)}$ has contributions from other orthogonal to the 1st root higher states, (its energy is close to the 3rd).

To show the (overall) behavior of $F_1$, in optimizing the energy of the 8-term 2nd HUM root (Fig. 4), each time the corresponding $F_1$ values, using the aforesaid fixed 1-term $\phi_0$, are additionally computed, differing near the minimum from the corresponding energies by $O(10^{-7})$. $F_1$ has minimum at $z_{1,1} = 1.44232$, while the difference $(F_1 - E_1)$ has minimum $O(10^{-9})$ at $z_{1,1} = 1.40329$ (instead of 1.44232 - indistinguishable in the figure), indicating that the 2nd HUM root $1^{(1,1,1,1,1)}$ is **indeed not a** Hamiltonian **eigenfunction** (that would have $F_1 - E_1 = 0$ ***at*** the minimum 1.44); the overlap with the above established $\psi_1 = 1^{(1,2,2,2,2)}$ is $\langle \psi_1 | 1^{(1,1,1,1,1)} \rangle = 0.875$. In lack of other parameters, by using $F_1$ the optimized "$F$" root $1^{(1,1,1,1,1)F}$ (tSzaR = $\Phi 100F$) nearly coincides with the 2nd HUM root near the minimum.

In a richer parameter space, including $z_{1,2}$ and, optionally, $a_{1,2,1} \neq 0$ (state-specific description), the optimized 8-term functions are as follows:

The optimized 8-term HUM 1st root $0^{(1,2,1,1,1)}$ has (tSzaR = $\Phi 010H$) E = −2.903123 a.u, $z_{0,1} = 1.93393$, $z_{0,2} = 3.50164$, {c} = {1, 0.290154, 0.184912, −0.077204, 0.006390, 0.014435, 0.017008, 0.011521}, $N = 1/2.94444$, with 2nd root deteriorated (tSzaR = $\Phi 0101$), E = −2.0196 a.u.. Thus, the 8-term $0^{(1,2,1,1,1)}$ is not essentially improved over the simpler 8-term $0^{(1,1,1,1,1)}$ above.

However, the optimized 8-term HUM 2nd root $1^{(1,2,1,1,1)}$ has (tSzaR = $\Phi 110H$) E = −2.14449 a.u., $z_{1,1} = 0.851359$, $z_{1,2} = 3.73405$ (or $z_{1,1} = 1.86703$, $z_{1,2} = 1.70272$), {c} = {1, 0.180637, −0.127452, 0.019460, −0.367640, −0.065135, −0.046299, −0.003814}, $N = 1/3.19489$, with 1st root deteriorated (tSzaR = $\Phi 1100$), E = −2.8886 a.u.. The overlap $\langle 0^{(1,2,1,1,1)} | 1^{(1,2,1,1,1)} \rangle = 0.0026 = (\Phi 010H | \Phi 110H)$.

By minimizing $F_1$ (using the above fixed 1-term $\phi_0$) first with $a_{1,2,1} = 0$, the same function is obtained, because the 1st root lies below $\phi_0$ - and the lowest root above $E[\phi_0]$ is the 2nd root. [It is not necessary to use any lower (fixed) $\phi_0$ because, if $a_{1,2,1} = 0$ there are no other parameters to vary, and, as seen in the next subsection, with $a_{1,2,1} \neq 0$, a minimum of $F_1$ is obtained with the same $\phi_0$ above.]

Allowing $a_{1,2,1} \neq 0$, the optimized 8-term HUM 2nd root $1^{(1,2,1,1,1)a}$ remains essentially the same: (tSzaR = $\Phi 111H$) E = −2.14449, $z_{1,1} = 1.86703$, $z_{1,2} = 1.70272$, $a_{1,2,1} = 1.5773 \times 10^{-7}$. The main orbitals resemble *1s1s′* (cf. Fig. 1(b) and Fig. 6). The overlaps with the above established $\psi_0 = 0^{(1,2,2,2,2)}$ and $\psi_1 = 1^{(1,2,2,2,2)}$ are: $\langle \psi_0 | 1^{(1,2,1,1,1)} \rangle = 0.0033$, and $\langle \psi_1 | 1^{(1,2,1,1,1)} \rangle = 0.9986$.

Generally, in a complete space both HUM and $F$ should yield the same wave function, but in a truncated space the 2nd HUM root may be worse than the lowest root above the known (highest) $E[\phi_{n-1}]$. Indeed, in 8-term $He_2$ $^1S$, at the $F_1$ minimum, the lowest three roots above the known 1-term $E[\phi_0] = -2.824$, are {−2.145, −2.028, −1.898}, whereas in optimizing the 2nd HUM root, the lowest three HUM roots are {−2.889, −2.144, −1.993} (As exposed later in detail, $F_1$ yields $E[\phi_1] = -2.1452$, much closer to the exact −2.1459 than the 2nd HUM root $E[\phi_1^{2r}] = -2.144$, where the 1st HUM root, $E[\phi_1^{1r}] = -2.889$, is much deteriorated).

Obviously, fulfilling the inherent restrictions of the HUM functions [cf. 5 and Eq. 1], the 8-term 2$^{nd}$ HUM root $1^{(1,1,1,1,1)}$ has lower quality than $0^{(1,1,1,1,1)}$ ($\approx 0^{(1,2,1,1,1)}$) since $0.875^2 < (1 − (1.3 \; 10^{-5})^2)$, but also the optimized HUM 2$^{nd}$ root $1^{(1,2,1,1,1)a}$ has lower quality as well, since $0.9986^2 < 1− (1.3 \; 10^{-5})^2$ - assuming near orthogonality (0.0026) to the optimized HUM 1$^{st}$ root.

*The main orbitals: $F_1$ (and OO) give $1s2s$; HUM gives $1s1s'$*

In contrast, by minimizing $F_1$, with the same set of parameters, ($z_{1,1}$, $z_{1,2}$, $a_{1,2,1}$), a much better 8-term approximant $1^{(1,2,1,1,1)F,a}$ is obtained: (tSzaR = Φ111F) $E = −2.145152$, $F = −2.145151$, $F – E = 5.3 \times 10^{-7}$, $z_{1,1} = 1.938718$, $z_{1,2} = 1.817736$, $a_{1,2,1} = 0.799286$, $\{c\} = \{1., 0.196506, 0.086316, −0.026204, 0.043226, 0.016366, 0.015810, 0.002479\}$, $N = 1/3.637910$, **where the main orbitals resemble $1s2s$** [cf. Fig. 1(c), and Fig.6 ]. The overlap with the optimized 8-term HUM 1$^{st}$ root is $\left\langle 0^{(1,2,1,1,1)} \middle| 1^{(1,2,1,1,1)F,a} \right\rangle = 0.0186 = (\Phi 010H|\Phi 111F)$. The corresponding overlaps with the above established basis are: $\left\langle \psi_0 \middle| 1^{(1,2,1,1,1)F,a} \right\rangle = 0.00490$, and $\left\langle \psi_1 \middle| 1^{(1,2,1,1,1)F,a} \right\rangle = 0.999807$. Clearly, the $F_1$-minimizing function, although truncated (8 terms), is much closer to the exact. This is also depicted in Fig. 3, where *it satisfies the Schrödinger equation better* ($−0.0003 <$ Schr $< 0.0009$) *than the 8-term HUM 2$^{nd}$ root* ($0.0006 <$ Schr $< 0.0012$).

Observe that contrary to the HUM 2$^{nd}$ root, which, despite its incorrect main orbitals $1s1s'$, is "corrected" by the $c$-series, $F_1$ directly, finds physically correct main orbitals $1s2s$ (and better energy in truncated space). It could be speculated that if this phenomenon emerges already in a system of two-electrons, it could be grosser (and more important) in large systems, where small truncated space is unavoidable and where the nature of the main (HOMO/LUMO) orbitals is decisive. This subject is previously unreported and is open to investigation. The above energies and overlaps are summarized in TABLE 1.

*Fulfillment of the three criteria by $F_1$ and Immediate improvement of $\Phi_0$.*

The functions obtained by $F_1$, fulfill the three **criteria** mentioned in the introduction:

First, to check the 27-term approximant $\psi_1$, a 2x2 ($H_{ij}−ES_{ij}$) generalized diagonalization between $\psi_1$ and the 1-term $\Phi_0$ indeed leaves $\psi_1$ practically unaffected, as "high 2x2 root" $\varphi_1 = (\psi_1 − 4.7 \; 10^{-5} \; \Phi_0)$, with energy changed by only $1.57 \; 10^{-9}$ a.u, while, by opening their "gap", improves $\Phi_0$ to the "low 2x2 root" $\varphi_0 = (0.999 \; \Phi_0 − 0.048 \; \psi_1)$, with (indeed lower than $E[\Phi_0] = −2.84766$) energy: $E[\varphi_0] = −2.84926$ a.u. differing from the energy of *the exact orthogonal to $\psi_1$* (in the $\{\Phi_0, \psi_1\}$ subspace) by $−1.57 \; 10^{-9}$ a.u. (which means that the "low 2x2 root" $\varphi_0$ is actually orthogonal to $\psi_1$). This verifies that $\psi_1$ is indeed very close to the exact excited eigenfunction, and that $\varphi_0$, in order to further approach $\psi_0$, needs be rotated only orthogonally to $\psi_1$. Similarly, a 2x2 diagonalization between the $F_1$ 27-term $\Phi_1$ and the 1-term $\Phi_0$ yields: $\varphi_1 = (\psi_1 − 1.4 \; 10^{-4} \; \Phi_0)$, $E[\varphi_1] – E[\Phi_1] = −1.34 \; 10^{-8}$ a.u., $\varphi_0 = (0.999 \; \Phi_0 − 0.048 \; \psi_1)$, with lower energy: $E[\varphi_0] = −2.84934$ a.u. differing from the energy of *the exact orthogonal to $\Phi_1$* (in the $\{\Phi_0, \Phi_1\}$ subspace) by $−1.34 \; 10^{-8}$ a.u., which verifies that the $F_1$ 27-term $\Phi_1$ is also very close to the exact. Now, to check $F_1$, a 2x2 diagonalization between the 8-term $\Phi_1$ and the 1-term $\Phi_0$ indeed leaves $\Phi_1$ also almost unaffected, as "high root" $\varphi_1 = (\psi_1 − 6.2 \; 10^{-4} \; \Phi_0)$, $E[\varphi_1] – E[\Phi_1] = −2.69 \; 10^{-7}$ a.u., and, by opening their "gap", improves $\Phi_0$ to the "low root" $\varphi_0 = (0.999 \; \Phi_0 − 0.052 \; \psi_1)$, with lower energy: $E[\varphi_0] = −2.84957$ a.u. differing from the energy of *the exact orthogonal to $\Phi_1$*

(in the {$\Phi_0$, $\Phi_1$} subspace) by $-2.69 \cdot 10^{-7}$ a.u., which means that the "low 2x2 root" $\varphi_0$ is practically orthogonal to $\Phi_1$ and verifies that $\Phi_1$ $1s2s$ is indeed very close to the exact excited eigenfunction; $\varphi_0$ could still be further improved by rotating it orthogonally to $\Phi_1$. Repeating the check with the 8-term HUM 2$^{nd}$ root $1s1s'$ and the 1-term $\Phi_0$, yields: $\varphi_1 = (0.999 \ \psi_1 - 1.2 \cdot 10^{-3} \ \Phi_0)$, $E[\varphi_1] - E[\Phi_1] = -1.02 \cdot 10^{-6}$ a.u., $\varphi_0 = (0.999 \ \Phi_0 - 0.049 \ \psi_1)$, with lower energy: $E[\varphi_0] = -2.84937$ a.u. differing from the energy of *the exact orthogonal to* $\Phi_1$ (in the {$\Phi_0$, $\Phi_1$} subspace) by $-1.02 \cdot 10^{-6}$ a.u., which verifies that 8-term HUM 2$^{nd}$ root $1s1s'$ is indeed inferior than the $F_1$ 8-term $1s2s$ function.

Secondly, as shown in Fig. 5, $F_1$ has local minimum, always above the energy, while (thirdly) the energy is a saddle point at the minimum of $F_1$, showing that it has approached the exact excited state $1s2s$. For some parameters ($z$, $c$, etc), the energy has local minimum, while for others (as $a_{1,2}$) it has local maximum.

*Quick check of reasonableness via the main orbitals*

Fig. 6 shows the main orbitals of both the 27-term and the 8-term optimized wave functions of the HUM 2$^{nd}$ root and also of $\phi_1^{LE}$ (the lowest orthogonal to $\phi_0$) and of $F_1$ excited state solutions, compared to the ground state. Observe that the HUM answer is $1s1s'$, instead of $1s2s$, (because $a_{1,2} \approx 0$, as mentioned above) and the HUM 8-term function deviates from the 27-term $\psi_1$, as anticipated in the introduction, (also the HUM $1s$ orbital is, unexpectedly, slightly more *diffuse* than $1s^2$), therefore the HUM function is veered away from the exact, whereas the $F_1$ answer is indeed $1s2s$, the $F_1$ 8-term function already almost coincides with the exact, and the $F_1$ $1s$ orbital is indeed more *compact* than $1s^2$, pushed toward the nucleus by the $2s$ electron, as intuitively expected.

$F_1$ allows one to quickly decide about the quality of the *main* orbitals, instead of "inspecting" them (that would be difficult to decide in large systems). If the function of the main orbitals is close to the excited eigenfunction, then $F_1$ will behave "reasonably" in varying one parameter about its final value, i.e. it will have a minimum above a critical point (minimum or a saddle point) of the energy. Fig. 7 shows that $F_1$ behaves "reasonably" around the $F_1$ $1s2s$ 8-term *main* orbitals (i.e. without the series expansion), but neither around the HUM $1s1s'$ 8-term main orbitals (again without the series expansion), nor around the 27-term (OO) lowest orthogonal to $\phi_0$ (again without the series expansion).

*The energy of $\phi_1^+$ (the closest to $\phi_1$ orthogonal to $\phi_0$ )*

Now, by using the above established $\Phi_1^{2r}$ as $\psi_1$ and a truncated approximant of $\psi_0$, as $\phi_0$, it is demonstrated that the corresponding $|\phi_1^+\rangle$ [cf. Eq. 2], i.e. the closest to $\psi_1$ while orthogonal to $\phi_0$, lies lower than $E[\psi_1]$ (actually: $E[\Phi_1^{2r}]$). Since the 8-term HUM 1$^{st}$ root $0^{(1,1,1,1,1)}$ is, as seen above, "almost perfect", $\langle\phi_0|1\rangle^2 = 10^{-10}$, $E[\phi_1^+]$ is lower than $E_1 = -2.14584$ by only $10^{-10}$ a.u., well beyond the accuracy of the present demonstration, so, a less accurate $\phi_0$ will be used in the demonstration, e.g. the 1-term "fixed", used in $F_1$, with $\langle\phi_0|1\rangle = 0.04786$; then by replacing $\psi_1$ by $\Phi_1^{2r}$ in $|\phi_1^+\rangle$: $\langle\phi_1^+|H|\phi_1^+\rangle = -2.15470$ a.u., whereas by replacing it in the exact formula [cf. Eq. 2]: $E[\phi_1^+] = -2.14745$ a.u.. The 0.3% discrepancy is due to the inexactness of $\Phi_1^{2r}$. We expect that if we further minimized the energy while keeping orthogonality to this

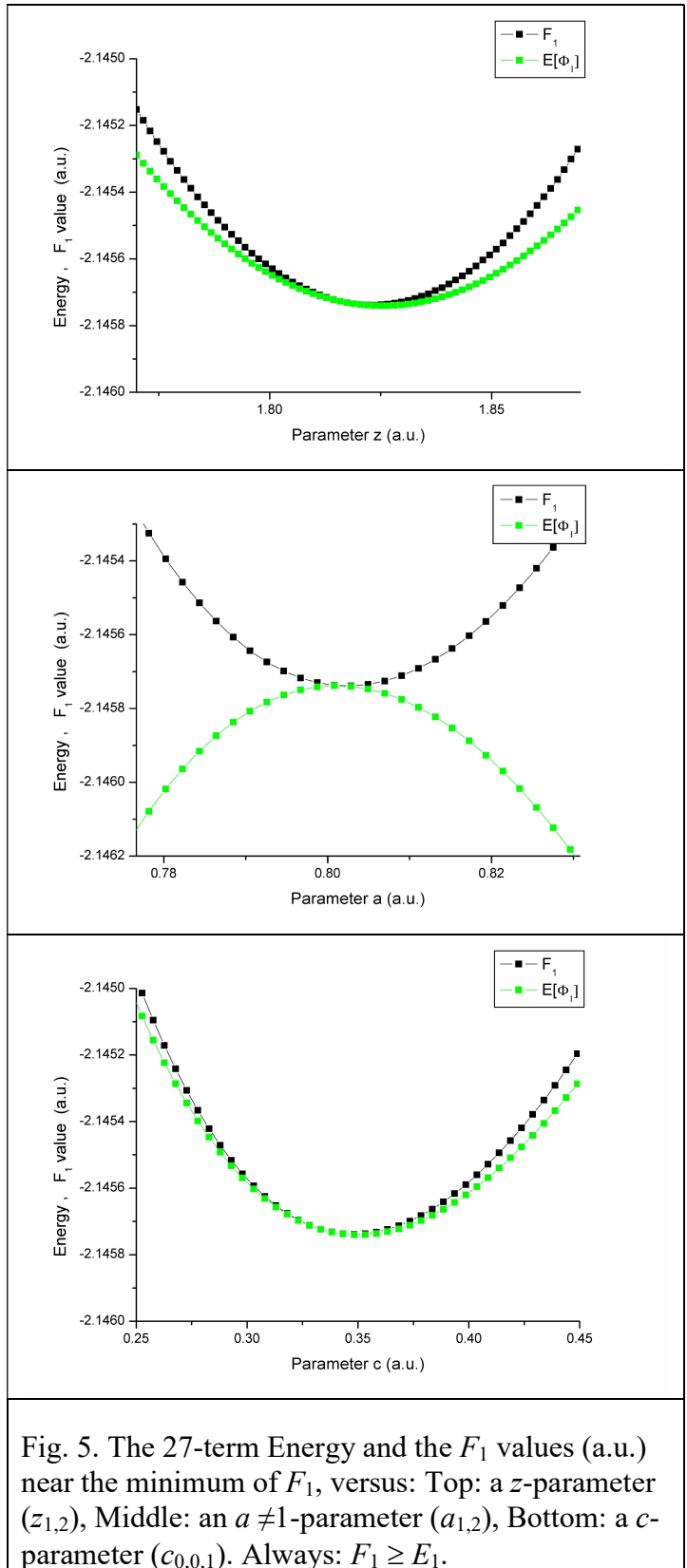

Fig. 5. The 27-term Energy and the $F_1$ values (a.u.) near the minimum of $F_1$, versus: Top: a $z$-parameter ($z_{1,2}$), Middle: an $a \neq 1$-parameter ($a_{1,2}$), Bottom: a $c$-parameter ($c_{0,0,1}$). Always: $F_1 \geq E_1$.

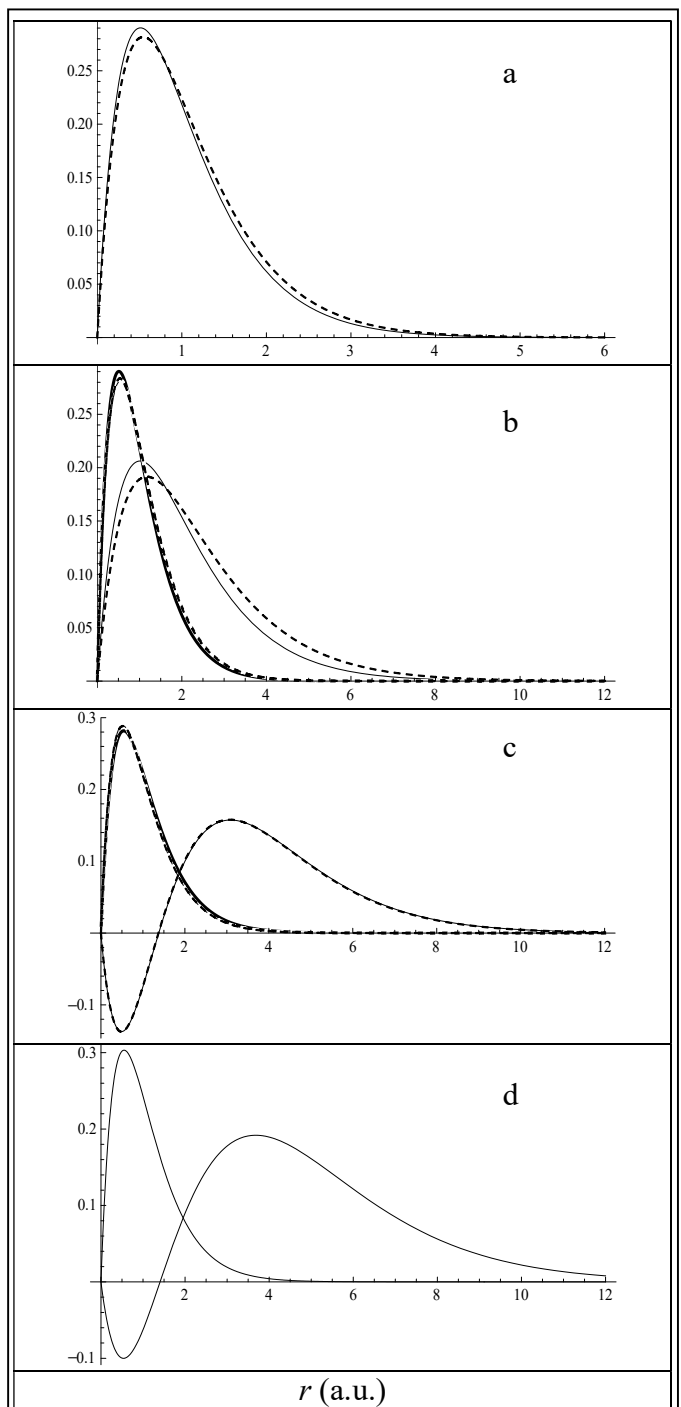

Fig. 6. Main orbitals of the 27-term (solid line) and 8-term (dashed line) optimal wave functions. a: Ground state $1s^2$. b: HUM excited state $1s1s'$ (either $a = 0$, or $a \approx 0$ as explained in the text, compared to $1s^2$. c: $F_1$ excited state $1s2s$ compared to $1s^2$. d: Lowest orthogonal to $\phi_0$.

1-term $\phi_0 = 0^{(0,1,0,0,0)}$, we should end up with a function orthogonal to this $\phi_0$, lying lower than −2.15470 a.u. (if this were the correct value based on the exact $\psi_1$), veered well away from $\psi_1$. This is demonstrated below. If the deteriorated HUM 1$^{st}$ root, orthogonal to the optimized HUM 2$^{nd}$ root $1^{(12111)a=0}$, is used as $\phi_0$, with $\langle \phi_0 | 1 \rangle$ = −0.00405, then the corresponding $\langle \phi_1^+ | H | \phi_1^+ \rangle$ = −2.14591 a.u. (and by the exact formula of Eq. 2: $E[\phi_1^+]$ = −2.14585 a.u.), which is slightly below $E[\Phi_1^{2r}]$, while the exact eigenvalue is slightly even lower, which confirms that it is not necessary for the 1$^{st}$ root to approach $\psi_0$, $\phi_0 \to \psi_0$, although deteriorated, only orthogonality to $\psi_1$, $\langle \phi_0 | \psi_1 \rangle \to 0$, is adequate.

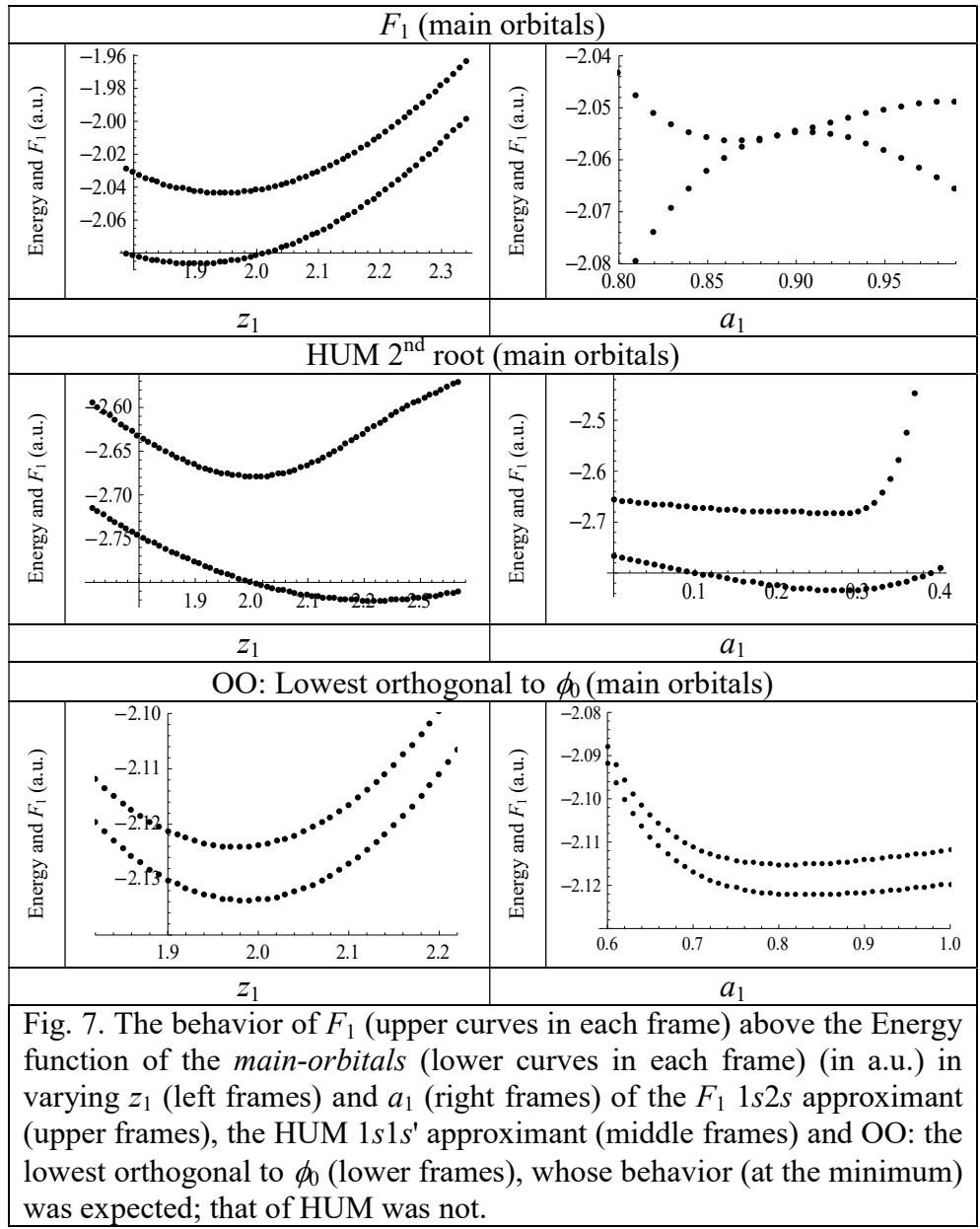

Fig. 7. The behavior of $F_1$ (upper curves in each frame) above the Energy function of the *main-orbitals* (lower curves in each frame) (in a.u.) in varying $z_1$ (left frames) and $a_1$ (right frames) of the $F_1$ 1s2s approximant (upper frames), the HUM 1s1s' approximant (middle frames) and OO: the lowest orthogonal to $\phi_0$ (lower frames), whose behavior (at the minimum) was expected; that of HUM was not.

*The lowest orthogonal to $\phi_0$ (OO)*

    Up to now it has been demonstrated that the optimized HUM 2$^{nd}$ root is veered away from the exact because it must be orthogonal to a deteriorated approximant of

the ground state while lying higher than the exact (in accordance with the HUM theorem). Next it will be demonstrated that minimizing the excited state energy while keeping orthogonality to a normalized *approximant* of the ground state, $\phi_0$, (e.g. the 1-term "fixed", used in $F_1$) leads to a function, $\phi_1^{LE}$, lying lower than the exact, therefore, also veered away from the exact. Thus the $F_1$-minimizing function, $\phi_1^{F_1}$, is the most reliable truncated approximant among the three $\{\phi_1^{F_1}, \phi_1^{LE}, \text{ and } \Phi_1^{2r}\}$:

From a trial function $\phi$ (un-normalized) subtract its projection to (normalized) $\phi_0$, to obtain the required normalized orthogonal, and its energy:

$$\phi_\perp \equiv \Phi_1 = \frac{\phi - \phi_0 \langle \phi_0 | \phi \rangle}{\sqrt{\langle \phi | \phi \rangle - |\langle \phi_0 | \phi \rangle|^2}} \quad ; \quad \langle \Phi_1 | H | \Phi_1 \rangle = \frac{\langle \phi | H | \phi \rangle - 2 \langle \phi_0 | H | \phi \rangle \langle \phi_0 | \phi \rangle + \langle \phi_0 | H | \phi_0 \rangle |\langle \phi_0 | \phi \rangle|^2}{\langle \phi | \phi \rangle - |\langle \phi_0 | \phi \rangle|^2} \quad ,$$

that will be used as a Rayleigh-Ritz quotient.

To obtain the energy minimum, either all parameters ($\{z\},\{a\},\{c\text{'s}\}$) of the trial function $\phi$ can be varied, or only the non-linear, ($\{z\},\{a\}$), be varied, while (for every trial $\{z\},\{a\}$) the linear part, $\{c\}$, can be reduced to a generalized eigenvalue problem $(\mathbf{A} - \lambda \mathbf{B}) \cdot \mathbf{c} = 0$, where $\mathbf{A}_{ij}$ are the coefficients of $c_i c_j$ of the numerator and $\mathbf{B}_{ij}$ of the denominator of the Rayleigh-Ritz quotient, and where $\lambda$ is a Lagrange multiplier, whose lowest value (1$^{st}$ root) is minimized by varying ($\{z\},\{a\}$). The corresponding lowest-lying eigenvector $\{c\}$ provides the coefficients of $\phi$, which, along with ($\{z\},\{a\}$), minimize $\langle \Phi_1 | H | \Phi_1 \rangle$, to obtain the final $\phi_1^{LE}$, by normalizing $\phi - \phi_0 \langle \phi_0 | \phi \rangle$.

After minimization of the above Rayleigh-Ritz quotient, $\phi_1^{LE}$ (without using $\psi_1$ or $\phi_1^+$) has energy $-2.14762$ a.u., certainly below the exact or of $\psi_1$, $-2.14584$. The 27-term minimizing function $\phi$, forms $\phi - \phi_0 \langle \phi_0 | \phi \rangle$, whose main orbital part has the form $(\upsilon(r_1)\chi(r_2) + \upsilon(r_2)\chi(r_1) - 1)\omega(r_1)\omega(r_2)$ where $\omega(r)$ is the 1$s$ orbital of $\phi_0$ (after normalization) and $\upsilon(r)$ and $\chi(r)$ are shown in Fig. 7. They remind 1$s$ and 2$s$ (where the 2$s$ is more remote - thus with lower energy), which means that $\phi_1^{LE}$, despite its lower energy, is not collapsed, but, of course, is veered away from the exact.

Thus, it has been demonstrated that, indeed, the $F_1$-minimizing function, $\phi_1^{F_1}$, is the most reliable truncated approximant among the three $\{\phi_1^{F_1}, \phi_1^{LE}, \text{ and } \Phi_1^{2r}\}$. It should be mentioned that if $\langle 1 | \phi_0 \rangle = 0$, these three functions coincide with $\psi_1$ and $\phi_1^+$, because the orthogonal to $\phi_0$ subspace will contain $\psi_1$, hence $\phi_1^+$ will coincide with $\psi_1$: not lying lower. Thus the minimizing function $\phi_1^{LE}$ will coincide with $\phi_1^+$ and, therefore, with $\psi_1$. Also, $\Phi_1^{2r}$, having $\phi_0$ as 1$^{st}$ root, will belong to the orthogonal subspace that contains $\psi_1$, and, since there is no other minimum than that of $\phi_1^{LE} = \psi_1$, it will coincide with $\psi_1$. Therefore, for the three functions to coincide the condition $\langle 1 | \phi_0 \rangle \to 0$ suffices; it is not necessary to meet the condition $\phi_0 \to \psi_0$.

Note that finding the minimum of $F_n$ is necessary: Fig. 8 shows all computed 8-term energy values, $E[\phi_1]$, in this work along their $F_1$ values, sorted by $F$. Of course, the first points have large $F$ and low $E$, and the last points tend to the minimum of $F_1$. Nevertheless, some of the last points, as shown in the inset, have reasonable $F_1$, but rather low $E[\phi_1]$. The correct point is the last one at the minimum of $F_1$, where the $E$ and $F$ values coincide. However, since $E_n$ is a saddle point, some wave functions $\phi_n$ near the $F_n$ minimum are still acceptable even if they lye below $E_n$, as long as they lye above $E_n - L$ [cf. Introduction/Recipe/note ], i.e. above the convex combination of all

lower eigenvalues up to $E_n$, which is of the form $(1-\sum_i a_i)E_n + \sum_i a_i E_i$; if the

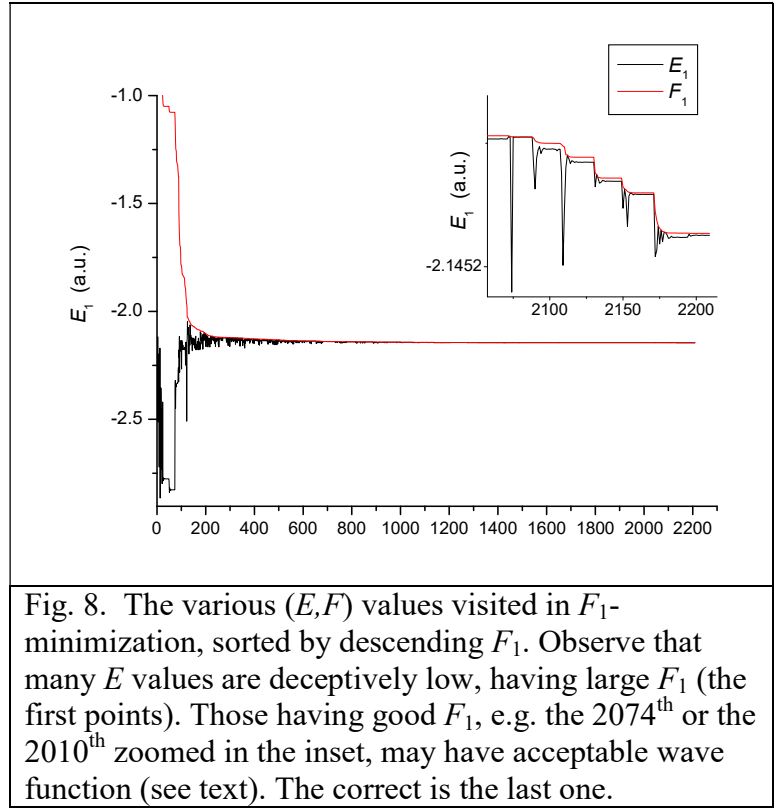

Fig. 8. The various $(E,F)$ values visited in $F_1$-minimization, sorted by descending $F_1$. Observe that many $E$ values are deceptively low, having large $F_1$ (the first points). Those having good $F_1$, e.g. the 2074[th] or the 2010[th] zoomed in the inset, may have acceptable wave function (see text). The correct is the last one.

expansion coefficients (weights $a_i = \langle i|\phi_n\rangle^2$) are small, or $L < \varepsilon$ (an accepted tolerance), the point is near $E_n$ and $\phi_n$ is acceptably near $\psi_n$. [45] The value of $L$, already at such accuracies near the $F_n$ minimum, can be easily checked. This criterion of validity holds for any $\phi_n$, not necessarily obtained via $F_n$. For example, $\phi_1^{LE}$, above, has $L = (E_1 - E_0)\langle 0|\phi_1^{LE}\rangle^2 = 0.0018$, which is a quite large tolerance, therefore, $\phi_1^{LE}$ is rather unacceptable.

**Demonstration of identifying a "flipped root"**

Let the ground and 1[st] excited wave functions of a hydrogen-like ion be parametrized as

$$\psi_0(z_0;r) = a_0(z_0)e^{-Z r z_0}, \quad \psi_1(z_1,g;r) = a_1(z_1,g)e^{-z_1 Z r/2}(1 - g Z r/2)$$

where Z is the nuclear charge, $z_0, z_1, g$ are variational parameters and $a_0(z_0)$, $a_1(z_1,g)$ are normalization constants. These functions are not orthonormal, unless $z_0 = 1$, $z_1 = 1$, $g = 1$, when they form eigenfunctions of the Hamiltonian $H\psi_j(r) = -\psi_j''(r)/2 - \psi_j'(r)/r - Z\psi_j(r)/r$.

In their 2x2 subspace create an orthonormal basis

$$\Psi_0(r) = \psi_0(r), \quad \Psi_1(r) = (\psi_1(r) - \psi_0(r)\langle\psi_0|\psi_1\rangle)/\sqrt{1 - \langle\psi_0|\psi_1\rangle^2},$$

whose overlap matrix is $\delta_{i,j}$. In diagonalizing their Hamiltonian matrix $H_{i,j} = \int_0^\infty 4\pi r^2 \, \Psi_i(r) H \Psi_j(r) \, dr$, let the two normalized eigenfunctions be $\Phi^{1r}(r)$, $\Phi^{2r}(r)$, with their eigenvalues ("roots") depending on $z_0, z_1, g$.

Now, around $z_0 \approx 2$ a root crossing occurs for a wide range around $z_1 \approx 1$ and $g \approx 1$ [cf. Fig. 9]. Near and "before" the crossing the continuation of $\psi_0(z_0;r)$ is

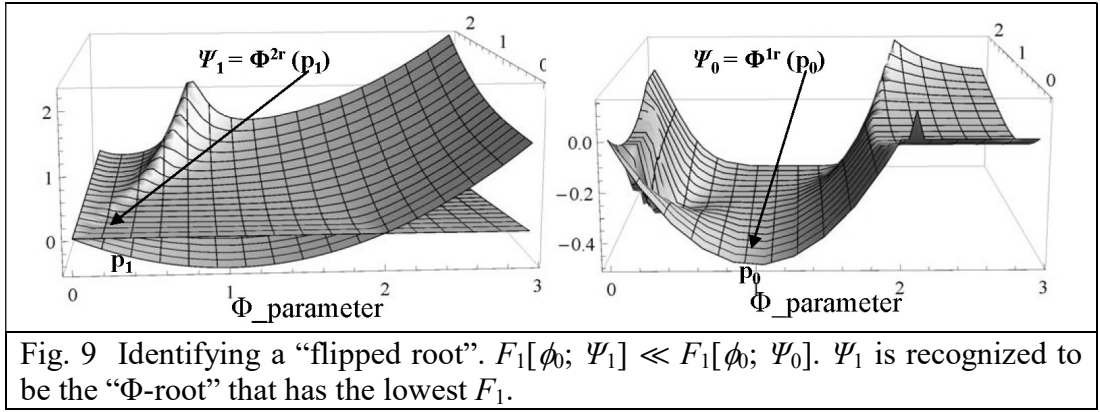

Fig. 9 Identifying a "flipped root". $F_1[\phi_0; \Psi_1] \ll F_1[\phi_0; \Psi_0]$. $\Psi_1$ is recognized to be the "$\Phi$-root" that has the lowest $F_1$.

$\Phi^{1r}(r)$ and the continuation of $\psi_1(z_1, g;r)$ is $\Phi^{2r}(r)$, whereas "beyond" the crossing the continuation of $\psi_1(z_1, g;r)$ is $\Phi^{1r}(r)$ and the continuation of $\psi_0(z_0;r)$ is $\Phi^{2r}(r)$. The question is to decide, via $F_1$, whether a given value of $(z_0, z_1, g)$, near the crossing, is "before" or "beyond" the crossing, in order to use the continuation, $\varepsilon$, of (always $\psi_1$:) $E[\psi_1(z_1, g;r)]$ in an optimization algorithm. (In the present demonstration Newton-Raphson (NR) is used: $\boldsymbol{\varepsilon'} := (\partial \varepsilon/\partial z_0, \partial \varepsilon/\partial z_1, \partial \varepsilon/\partial g) = 0$ is solved by proceeding iteratively to a new point $\mathbf{p} + \delta\mathbf{p} = \mathbf{p} - \mathbf{J}^{-1} \cdot \boldsymbol{\varepsilon'}$-or less if the method diverges- having started at some point $\mathbf{p} = (z_0, z_1, g)$, where $\mathbf{J}$ is the Jacobian matrix - Hessian of $\varepsilon$. [50])

Thus, consider

$$F_1[\phi_0; \Phi^{nr}] = E[\Phi^{nr}] + 2\frac{(\langle\phi_0|H|\Phi^{nr}\rangle - E[\Phi^{nr}]\langle\phi_0|\Phi^{nr}\rangle)^2}{(E[\Phi^{nr}] - E[\phi_0])(1 - \langle\phi_0|\Phi^{nr}\rangle^2)}, \quad n = 1, 2$$

and let a fixed predetermined (deliberately not very accurate) ground state approximant be $\phi_0(z_0^0;r) = a_0(z_0^0) e^{-Z r z_0^0}$ with $z_0^0 = 1 - 0.05$. (By direct minimization, $F_1[\phi_0; \Phi^{2r}]$ is minimized at $z_1 = 1, g = 1$, ($z_0 \approx 1$), giving $F_1 \to E[\Phi^{2r}(r)] = E[\psi_1(1,1;r)] = -0.125$.) As explained in the introduction, among the two "roots", the continuation of the excited state, near the crossing, is the one with the lowest $F_1$. Indeed, for $Z = 1$, using (first traditionally), always the "2$^{nd}$ root", i.e. keeping $\varepsilon$ to $E[\Phi^{2r}(r)]$, [regardless of which $n$ ($n^{th}$ root) the lowest $F_1$ suggests] "root-flipping" shows up: The sequence of TABLE 2 is obtained, even by using half NR-step, whereas, by consulting $F_1$ the continuation of the excited state is

recognized near the crossing and used (until finally, at convergence, only $n = 2$, the

| TABLE 2. Root flipping: No convergence of the 2$^{nd}$ root | | | | |
|---|---|---|---|---|
| | | | n | $E^{2r}$ |
| 2.7 | 0.95 | 0.8 | 1 | 1.09915 |
| 1.89975 | 0.657935 | 0.618574 | 1 | 0.0312309 |
| 1.79561 | 0.457167 | 0.5843 | 2 | -0.05816 |
| 1.81109 | 0.303372 | 0.471712 | 2 | -0.0491109 |
| 1.88811 | 0.0253053 | 0.0834495 | 2 | -0.0056792 |
| -4.64705 | 1.13597 | 3.82639 | 1 | 14.4125 |
| -4.64705 | 1.13597 | 3.82639 | 1 | 14.4125 |
| -1.79846 | 0.463128 | 1.53227 | 1 | 3.19065 |
| 0.416586 | -0.117902 | -0.379791 | 2 | 0.0138916 |
| 0.647726 | -0.168901 | -0.60433 | 2 | 0.0232773 |
| 0.913718 | -0.21854 | -0.88652 | 2 | 0.0336386 |
| 1.15883 | -0.254786 | -1.19969 | 2 | 0.0428655 |
| ... | ... | ... | ... | ... |

2$^{nd}$ root, is suggested by the lowest $F_1$) [cf. TABLE 3].

| TABLE 3. Same as Table 2, but by consulting $F_1$ | | | | |
|---|---|---|---|---|
| $z_0$ | $z_1$ | g | n | $\varepsilon$ |
| 2.7 | 0.95 | 0.8 | 1 | -0.142331 |
| 2.62754 | 0.95588 | 0.907362 | 1 | -0.127151 |
| 2.37234 | 0.971824 | 0.98181 | 1 | -0.124935 |
| 1.92938 | 0.992178 | 1.00683 | 1 | -0.125094 |
| 1.3233 | 1.02512 | 0.953813 | 2 | -0.124875 |
| 1.18358 | 1.01644 | 0.958059 | 2 | -0.12496 |
| 1.07193 | 1.01041 | 0.964645 | 2 | -0.124988 |
| 0.980565 | 1.00646 | 0.97107 | 2 | -0.124997 |
| 0.905827 | 1.00399 | 0.976612 | 2 | -0.124999 |
| 0.844753 | 1.00248 | 0.981153 | 2 | -0.125. |

Observe that at the beginning, "beyond" the crossing, the lowest $F_1$ dictates to use, for the next step, the (lower than $E_1$) value of $\varepsilon = E\left[\Phi^{1r}(r)\right]$ ($n = 1$, the lowest function at that point).

Similarly, using only the 2$^{nd}$ root (regardless of which $n$ is dictated by the lowest $F_1$) and starting, again "beyond" the crossing ($n = 1$) from the same point, TABLE 4 is obtained. In this case, despite the original irregularities due to root-flipping, the 2$^{nd}$ root finally remained "before" the crossing ($n = 2$), and converged. 0.3 of NR-step was used. Now, by consulting $F_1$ no irregularities occurred [cf. TABLE 5].

Note that finally, near the minimum of the 2$^{nd}$ root $E\left[\Psi_1(\mathbf{p}_1)\right] > E\left[\Psi_0(\mathbf{p}_1)\right]$, "before" the crossing, the convergence *should* use the 2$^{nd}$ root. If (and when), while in $n = 1$, it approached a point of the 1$^{st}$ root, i.e. "beyond" the crossing, the NR-step should be increased somewhat in the direction of the last "false"-converged δ$\mathbf{p}$ in order to send it farther to $n = 2$, i.e. "before" the crossing where the real minimum should be [cf. TABLE 6]. (At convergence this should always be checked). While in $n = 1$, it approached p = (2.34, 0.94, 1.01); in order to send it farther to $n = 2$ "before"

the crossing, p was extrapolated every 3 steps by twice the final step - or more if convergence is slow, according to the logical code exposed in TABLE 7. Such an extrapolation was also used in the first example above; in the second it was not needed.

| TABLE 4. Accidental convergence of the 2nd root without consulting $F_1$ | | | | |
|---|---|---|---|---|
| $z_0$ | $z_1$ | g | n | $E^{2r}$ |
| 2.7 | 1.2 | 1.1 | 1 | 0.99477 |
| 2.20242 | 0.940034 | 0.917342 | 1 | 0.256091 |
| 1.89495 | 0.731999 | 0.818135 | 1 | -0.0551044 |
| 1.58749 | 0.523964 | 0.718929 | 2 | -0.0866029 |
| -3.05737 | -10.9139 | -12.3444 | 1 | 69.042 |
| -2.07399 | -7.38161 | -8.30417 | 1 | 33.6488 |
| -1.38465 | -4.90769 | -5.47902 | 1 | 16.3024 |
| -0.695314 | -2.43378 | -2.65387 | 1 | 5.07481 |
| -0.695314 | -2.43378 | -2.65387 | 1 | 5.07481 |
| -0.414382 | -1.43604 | -1.53652 | 1 | 2.2856 |
| -0.214196 | -0.731385 | -0.76036 | 1 | 0.910672 |
| -0.0140099 | -0.0267279 | 0.0157957 | 2 | 0.0214185 |
| 0.164968 | 0.31467 | -0.18661 | 2 | -0.0598432 |
| 0.326748 | 0.626253 | -0.334901 | 2 | -0.103163 |
| 0.401263 | 0.771636 | -0.381191 | 2 | -0.116354 |
| ... | ... | ... | ... | ... |
| 0.518132 | 1.00162 | -0.429903 | 2 | -0.124998 |
| 0.516411 | 0.998281 | -0.42876 | 2 | -0.125. |

It should be pointed out (generally for fixed positions of the nuclei) that state averaging *at* the crossing in the parameter space is useless, because we want the functions at $\mathbf{p}_0$ (the ground state) and at $\mathbf{p}_1$ (the excited state), both before the crossing [cf. Fig. 9].

| TABLE 5. Same as in Table 3, but by consulting $F_1$ | | | | |
|---|---|---|---|---|
| $z_0$ | $z_1$ | g | n | ε |
| 2.7 | 1.2 | 1.1 | **1** | -0.119716 |
| 1.93961 | 1.03415 | 0.976157 | 1 | 0.126188 |
| 1.76764 | 0.886719 | 1.08156 | **2** | -0.121279 |
| 1.61807 | 0.892447 | 1.16453 | 2 | -0.121751 |
| 1.41702 | 0.918759 | 1.21697 | 2 | -0.123372 |
| 1.20892 | 0.9443 | 1.21744 | 2 | -0.124476 |
| 1.04199 | 0.963778 | 1.18615 | 2 | -0.12487 |
| 0.924215 | 0.977699 | 1.15139 | 2 | 0.124969 |
| 0.841179 | 0.986541 | 1.12222 | 2 | 0.124992 |
| 0.780975 | 0.991767 | 1.09888 | 2 | 0.124998 |
| 0.735948 | 0.994853 | 1.0804 | 2 | -0.124999 |
| 0.701294 | 0.996712 | 1.06577 | 2 | -0.125. |

Note that the graphs of all converged functions, above,
$$\Phi^{2r}(r) = -0.0074983\, e^{-0.844753\, r} + (0.206437 - 0.101273\, r)\, e^{-0.50124\, r},$$

$$\Phi^{2r}(r) = 0.615983\, e^{-0.516411\, r} - (0.416497 + 0.0892886\, r)\, e^{-0.499141\, r},$$

TABLE 6. Overpassing a false convergence

| $z_0$ | $z_1$ | g | n | E |
|---|---|---|---|---|
| 2.5 | 1.2 | 0.9 | *1* | -0.138096 |
| 2.62454 | 1.03587 | .935687 | 1 | -0.127460 |
| 2.48591 | .989487 | .976356 | 1 | -0.125154 |
| 2.34727 | .943106 | 1.01703 | *1* | -0.124859 |
| 1.41343 | 1.02160 | .963025 | *2* | -0.124868 |
| 1.25337 | 1.01623 | .964009 | 2 | -0.124954 |
| 1.12615 | 1.01049 | .968734 | 2 | -0.124986 |
| 1.02330 | 1.00655 | .974106 | 2 | -0.124996 |
| .939738 | 1.00405 | .978974 | 2 | -0.124999 |
| .871820 | 1.00250 | .983044 | 2 | -0.125. |

$$\Phi^{2r}(r) = 0.0178076\, e^{-0.701294\, r} + (0.182078 - 0.0970265\, r)\, e^{-0.498356\, r},$$

$$\Phi^{2r}(r) = -0.00706569\, e^{-0.87182\, r} + (0.205952 - 0.10123\, r)\, e^{-0.50125\, r},$$

are practically identical to

$$\psi_1(1,1;r) = (1 - r/2)\, e^{-r/2} / \sqrt{8\pi} = (0.199471 - 0.0997356\, r)\, e^{-r/2};$$

TABLE 7. Three-step NR extrapolation

```
if |p3-p2|>| p2-p1| then
        h=p3
else
        h=p3+(p3-p2),

if n(p1)≡1 then
        { p2=NR(p1) and
        if n(p2)≡1 then
                [ p3=NR(p2) and
                if n(p3)≡1 then
                        (p=h
                        )
                else if n(p3)≠1 then
                        p=NR(p3)
                ]
        else if n(p2)≠1 then
                p=NR(p2)
        }
else if n(p1)≠1 then
p=NR(p1)
```

they differ by at most 0.001 at $r = 0$ (the second differs at most by $1.5 \times 10^{-5}$) [cf. Fig. 10].

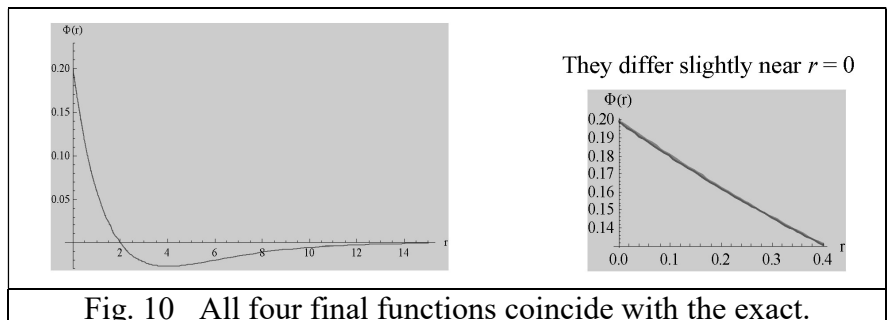

Fig. 10   All four final functions coincide with the exact.

**Application to conventional configuration interaction treatment**

The use of $F_n$ has been preliminarily applied to the computation of atomic excited states in standard coordinates $(r, \theta, \varphi)$ by configuration interaction (CI) using analytic Laguerre-type orbitals whose polynomial prefactors are variationally optimized (AVOLTOs), thus providing small and concise analytic wave functions, while achieving accuracy comparable to numerical MCSCF. [45]

The atomic wave function is a normalized CI sum whose configurations are formed out of Slater determinants (SD), composed of atomic spin-orbitals, whose spatial parts are

$$\langle r | n, \ell, m \rangle = A^{n,\ell,m} L_{n,\ell}(r) Y^{\ell,m}(\theta, \phi),$$

where $A^{n,\ell,m}$ are normalization factors, $Y^{\ell,m}(\theta,\phi)$ are spherical harmonics, and $L_{n,\ell}(r)$ are the AVOLTOs,

$$L_{n,\ell}(r) = \sum_{k=0}^{n-\ell-1} c_k^{n,\ell} g_k^{n,\ell} r^{(\ell+k)} e^{-z_{n,\ell} \frac{r}{n}} + b^{n,\ell} e^{-q^{n,\ell} z_{n,\ell} \frac{r}{n}} \delta_{\ell,0}. \tag{5}$$

Here $c_k^{n,\ell}$ are the usual associated Laguerre polynomial coefficients, $z_{n,\ell}$, $b^{n,\ell}$, $q^{n,\ell}$ are variational parameters and $g_k^{n,\ell}$ are factors determined by orthogonalization to desired, only, orbitals, i.e. by solving $\langle n_i, \ell, m | n_j, \ell, m \rangle = \delta_{i,j}$, for $g_k^{n,\ell}$, so that not all orbitals are mutually orthogonal. Therefore the general formalism is non-orthogonal, allowing spin unrestricted, as well as open shell computation.

Thus, given the atom with nuclear charge $Z_{nuc}$, and $N$ electrons, with space and spin coordinates $\mathbf{r}_1 s_1$, ..., $\mathbf{r}_N s_N$, as well as the symmetry type and the electron occupancy, the desired $N$-electron normalized wave function, for the $n^{th}$ excited state, consisting of $N_{conf}$ (predetermined) configurations, out of $N_{det}$ SDs, is

$$\phi_n(\mathbf{r}_1 s_1, ..., \mathbf{r}_N s_N) = \sum_{p=1}^{N_{conf}} d_p \sum_{a=1}^{N_{det}} f_{p,a} D_a \; ; \; |\phi_n|^2 = 1, \tag{6}$$

obtained by minimimization of $F_n$, with Hamiltonian

$$H = -\frac{1}{2} \sum_{i=1}^{N} \left( \nabla_i^2 + \frac{Z_{nuc}}{|r_i|} \right) + \sum_{i>j}^{N} \frac{1}{|r_i - r_j|} \equiv \sum_{i=1}^{N} h_i + \sum_{i>j}^{N} g_{i,j}.$$

The $D_\alpha$ are all ($N_{det}$) (consistent with the desired electronic state) SDs, formed out of $N_{orb}$ (predetermined), spinorbitals $a_i$, - to be optimized - and $f_{p,a}$ are all ($N_{conf}$ x $N_{det}$) consistent corresponding coefficients. The linear parameters $d_p$ are determined from a desired root of the secular equation ($N_{conf}$ x $N_{conf}$) with $(p,q)$ matrix elements

$$\sum_{a,b=1}^{N_{det}} f_{p,a} f_{q,b} \langle D_a | H - E | D_b \rangle,$$

where $\phi_n$ is to approach a critical point of both the energy and $F_n$, while all $\phi_i$ $i<n$, remain unvaried rather crude approximants of the lower states.

The one- and two-electron terms between SDs are computed by

$$\langle D_a | \sum_{i=1}^{N} h_i | D_b \rangle = \frac{1}{\sqrt{D_{aa} D_{bb}}} \sum_{i,j=1}^{N} \langle a_i | h | b_j \rangle D_{ab}(a_i b_j)$$

where $a_i, b_j, \ldots$ are the spin-orbitals, $D_{ab} = \det |\langle a_1|b_1\rangle \langle a_2|b_2\rangle \cdots \langle a_n|b_n\rangle|$, and $D_{ab}(a_i b_j)$ denotes the cofactor of the element $\langle a_i|b_j\rangle$ in the determinant $D_{ab}$, and $D_{aa}$, $D_{bb}$ are similar normalization factors. Also,

$$\langle D_a | \sum_{i>j}^{N} g_{i,j} | D_b \rangle = \frac{1}{\sqrt{D_{aa} D_{bb}}} \sum_{j>l}^{N} \sum_{i>t}^{N} \langle a_i a_t | g | b_j b_l \rangle D_{ab}(a_i a_t b_j b_l)$$

where $D_{ab}(a_i a_t b_j b_l)$ is the cofactor of $D_{ab}$ defined by deleting the rows and columns containing $\langle a_i|b_j\rangle$ and $\langle a_t|b_l\rangle$ and attaching a factor $(-1)^{i+j+t+l}$ to the resultant minor. The determinant cofactors are most efficiently computed via the inverse matrix, after adding to all matrix elements small random numbers of the order of machine accuracy, in order to avoid occasional (but harmless) vanishing of the determinant.

For the 1$^{st}$ excited state of He $^1S$ 1s2s, using, in $\phi_1$, 11 AVOLTOs forming 76 SDs and 22 configurations, and a fixed crude $\phi_0$ approximant of 1s$^2$ of only two AVOLTOs in a 2x2 CI (E[$\phi_0$] = −2.88 a.u.), then E[$\phi_1$] = −2.1458140 a.u. ($F_1$ = −2.1458139 a.u.). The wave function is primarily $\phi_1$ = 0.9993 (1sB 2sA − 1sA 2sB) + 0.0190 (2p$_1$A 3p$_{-1}$B − 2p$_1$B 3p$_{-1}$A − 2p$_0$A 3p$_0$B + 2p$_0$B 3p$_0$A + 2p$_{-1}$A 3p$_1$B − 2p$_{-1}$B 3p$_1$A) + 0.0178 3s$^2$, where $A,B$ denote the spin, the {$r_{rms}$ distance from the nucleus (a.u); $z$, $b$, $q$} values are, for 1s: {0.8699666 ; 1.9690983, 0.0881525, 1.2060984}, for 2s: {5.6429592 ; 1.1248561, 0, 1}, for 2p: {1.1438142 ; 4.7885604, 0, 1} and for 3s: {1.0879814 ; 5.6155136, 0, 1}, the $g_k$-factors are, for 2s: (0.8186816, 1) and for 3s: (4.0287651, 2.0322568, 1), making them both orthogonal to another 1s, a little more diffuse: {1.0374887 ; 3.0931600, −0.9341100, 0.9519696}. This energy value can be compared to: (i) −2.1457316 a.u., with 10 AVOLTOs forming 68 SDs and 18 configurations, exactly orthogonal to a quite accurate $\phi_0$ of E = −2.9031501 a.u. with 15 AVOLTOs, 157 SDs and 40 configurations, and to (ii) −2.145873 a.u. using 10 Numerical MCHF orbitals, with a comparably highly accurate ground state of E = −2.9031173 E$_h$. [51]

For the 2$^{nd}$ excited state of He $^1S$ 1s3s, using, in $\phi_2$, 11 AVOLTOs, 77 SDs, 23 configurations, and again 2x2 $\phi_0$ and $\phi_1$, then E[$\phi_2$] = −2.0612263 a.u. ($F_1$ = −2.0611758 a.u.). The wave function is primarily $\phi_2$ = 0.9788 (1sA 3sB − 1sB 3sA) + 0.2035 (3sA 2sB − 3sB 2sA) + 0.0170 (1sA 2sB − 1sB 2sA), the {$r_{rms}$; $z$, $b$, $q$} values are, for 1s: {1.1051964 ; 1.61507, 2.0564, 0.95729}, for 3s: {12.9875477 ; 1.09458, 0, 1}, for 2s: {1.6490337 ; 3.49456, 0, 1}, the $g_k$-factors are, for 3s: (0.7660135, 0.9277702, 1), making it orthogonal to 1s and to the previous 2s, and for 2s: (1.5813926, 1), making it orthogonal to 1s. This energy value can be compared to −2.0612681, obtained by *B-splines*. [52] By increasing to 19 AVOLTOs, 263 SDs, 53 configurations, the improved energy is E[$\phi_2$] = −2.0612522 a.u. (= $F_1$) [7].

**Summary and conclusions**

The energy of any excited state is a saddle point in the parameter space of the wave function expansion.

In truncated expansions, the wave function obtained as an optimized higher root of the secular equation, according to HUM [3] theorem approaches the excited energy from above (Energy ≥ $E_{exact}$), but, the function itself is veered away from the exact eigenfunction because it is orthogonal to deteriorated lower roots, whose orthogonal subspace cannot contain the exact excited state. The truncated wave function is unable to approach the exact eigenfunction at any desired accuracy (it should better be avoided), and the usual remedy is to use tremendously large expansions, impractical for large systems, even if the excited state under description is rather simple. It has been demonstrated that for He $^1$S $1s2s$, the main orbitals of the 2$^{nd}$ HUM root are essentially $1s1s'$, and a large expansion is needed to fix it.

On the other hand, using truncated wave functions orthogonal to lower lying truncated approximants (OO), is not safe if orthogonality to the main only orbitals is secured, and if orthogonality to the whole lower approximants is used, then, due to the saddleness of the excited energy, the energy has to go below the exact, down to $E > E_{exact} − L$ [cf. Introduction/Recipe/(b)]. The truncated wave function is also veered away from the exact eigenfunction, unable to approach it, without necessarily being collapsed (this method should also be better avoided). For the same reason, if a function is close to the exact and $L$ is acceptably small, then despite the slightly lower energy, the function is acceptable.

The safest method to approach the exact eigenfunction in a truncated space is by minimizing the functional $F_n$, needing rather crude lower lying approximants (otherwise the method would be impractical): Independently of the accuracy of the lower lying approximants, $F_n$ at the minimum ($F_n \geq E_{exact}$) approaches the exact eigenfunction at any desired accuracy (i.e. there is no restriction, like HUM, to be deteriorated compared to the optimized root). It has been demonstrated that for He $^1$S $1s2s$, the main orbitals obtained by minimizing $F_n$, are indeed $1s2s$, and a large expansion is not needed to provide a comprehensible physical picture. Using Hylleraas coordinates and a crude 1-term approximant of the ground state in $F_1$, the 8-term series expansion is already very close to the 27-term function and to the exact. (As the size of the expansion increases, all of the three methods, HUM, OO, and $F_n$, tend to the exact eigenfunction.) The method has been also applied by conventional configuration interaction using variationally optimized (via $F_n$) analytic Laguerre-type orbitals, yielding concise and comprehensible wave functions of comparable accuracy to numerical MCSCF, to the first three lowest states of He $^1$S ($1s^2$, $1s2s$ and $1s3s$) with comparisons to the literature. The small and comprehensible wave function can be safely used to correctly compute other properties beyond the energy, like the mean values of $r$, $r^2$, $1/r$ [7].

**Benefits:**

The main benefit of the $F_n$ method is that no orthogonality to lower lying approximants has to be imposed: Orthogonality should be an outcome as an *a-posteriori* output test. Also, rather crude lower approximants are needed, reducing significantly the number of two-electron integrals in each minimization step. Neither 2$^{nd}$ derivatives are needed, nor any new Ritz quotient, beyond the Hamiltonian. The extra cost is to compute a few more integrals with the crude lower approximants. If they are unreasonably crude, $F_n$ will simply drop to $−\infty$, an indication to use a less crude lower approximant.

After the 1$^{st}$ excited state is found (independently of the ground state), $F_n$ allows immediate improvement of the ground state approximant by a simple 2x2 diagonalization between the two approximants, the ground and the excited. If the excited is accurate, it remains unchanged, and the ground is improved toward a function that is orthogonal to the excited.

A third benefit is that $F_n$ allows recognition of a flipped MCSCF root near an avoided crossing, in the parameter space, without needing to compute 2$^{nd}$ derivatives or state average. Using fixed crude lower approximants, $F_n$ is computed of the two ambiguous functions. The function that yields the lowest $F_n$ is the (or the continuation of the) excited state. Since the minimum of $F_n$ is away "before" the avoided crossing in the parameter space, then state averaging *at* the crossing in the parameter space is useless. Thus, $F_n$ can be easily used to guide any standard MCSCF computation, without resorting to state averaging.

Finally, as shown in the Appendix, if the lowest lying states are known, it is possible to combine any higher lying function with, for example $\psi_0$, to obtain a function $\Phi$ having the energy of $\psi_1$, $E[\Phi]=E_1$, and, simultaneously, $\Phi$ being orthogonal to $\psi_1$. Since entanglement of states is already experimentally accomplishable [53], such an engineered superposition, if accomplished, could in principle be used to manipulate a chemical reaction occurring at energy $E_1$ without using the state $\psi_1$, but rather (perhaps easier to use) more remote electrons.

**Appendix**

**A.** To leading order in coefficients, the overlap and the Hamiltonian matrix elements are

$$\langle \phi_i | \phi_n \rangle = \langle i | \phi_n \rangle + \langle n | \phi_i \rangle + \cdots$$
$$\langle \phi_i | H | \phi_n \rangle = E_i \langle i | \phi_n \rangle + E_n \langle n | \phi_i \rangle + \cdots$$

Substituting $\langle i | \phi_n \rangle$ to each term of $L = \sum_{i<n}(E_n - E_i)\langle i | \phi_n \rangle^2$ gives, to leading order, $\left( E_n \langle \phi_i | \phi_n \rangle - \langle \phi_i | H | \phi_n \rangle \right)^2 / (E_n - E_i)$, which suggests an examination, in terms of known quantities, of

$$\sum_{i<n} \frac{\left( E[\phi_n]\langle \phi_i | \phi_n \rangle - \langle \phi_i | H | \phi_n \rangle \right)^2}{E[\phi_n] - E[\phi_i]},$$

which, when both $|\phi_i\rangle = |i\rangle$ and $U = \sum_{i>n}(E_i - E_n)\langle i | \phi_n \rangle^2 \to 0$, as directly verified, reduces to

$$L\left(1 - \sum_{i<n} \langle \phi_i | \phi_n \rangle^2 \right),$$

therefore, for $U \neq 0$ the behaviour of the paraboloid $E + 2L$ close to $|\phi_n\rangle = |n\rangle$ is adequately described by the functional $F_n$ in Eq. 3, the adequacy being specified by Hessian determinants and its principal minors, Eq. 4.

**B.** In Fig A.1, **g**, **a**, and **e** indicate the exact (unknown) ground, 1$^{st}$ and 2$^{nd}$ excited states, $|\psi_0\rangle$, $|\psi_1\rangle$ and $|\psi_2\rangle$ respectively. **G** is a (known) ground state approximant $|\phi_0\rangle$. **E** is orthogonal to **a**($|\psi_1\rangle$) and **G**($|\phi_0\rangle$), close to **e**($|\psi_2\rangle$), therefore, lying *above*

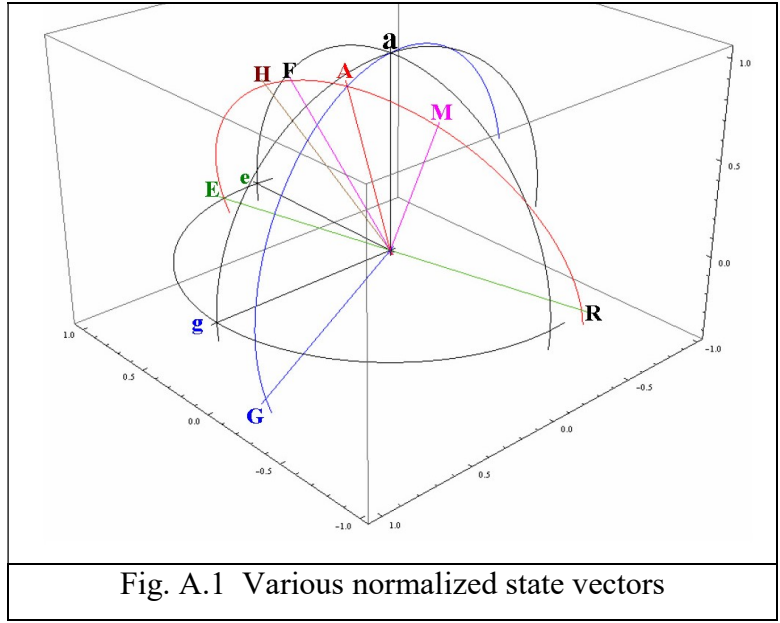

Fig. A.1 Various normalized state vectors

$E_1$. **A** is $|\phi_1^+\rangle$, orthogonal to **G**($|\phi_0\rangle$), and is the closest to **a**($|\psi_1\rangle$), and it lies *below* $E_1$ (see text). Therefore, on the orthogonal to **G**($|\phi_0\rangle$) subspace (circle **EA**), there is a state **F**, lying *at* $E_1$. $E[\mathbf{F}] = E_1 = E[|\psi_1\rangle]$. **F** is veered away from both **A**($|\phi_1^+\rangle$) and **a**($|\psi_1\rangle$). Optimizing the 2$^{nd}$ HUM root deteriorates the 1$^{st}$ HUM root (orthogonal to the 2$^{nd}$). Suppose, to avoid complicating the figure, and without loss of generality, that **G** is the deteriorated 1$^{st}$ HUM root. Then the optimized 2$^{nd}$ HUM root $|\phi_1^{2r}\rangle$ (**H**) belongs to the orthogonal to **G** circle **EFA**, and lies *above* (at least *at*) $E_1$, according to the HUM theorem. Therefore **H**($|\phi_1^{2r}\rangle$) **is more than the F veered away from both A($|\phi_1^+\rangle$) and a($|\psi_1\rangle$)**.

Since **A**($|\phi_1^+\rangle$), lying *below* $E_1$, is not an eigenfunction, then there is a state (**M**), orthogonal to **G**($|\phi_0\rangle$), that belongs to the orthogonal circle **EFA**, and lies *even lower* than (at most *at*) $E_1$, obtained by optimizing orthogonally to **G**($|\phi_0\rangle$) (OO in the text). Therefore **M**($|\phi_1^{OO}\rangle$) **is also veered away from both A($|\phi_1^+\rangle$) and a($|\psi_1\rangle$)**.

Since **M** is the minimum [orthogonal to **G**($|\phi_0\rangle$)] by continuing beyond **M** on the orthogonal circle **EFA**, the value $E_1$ can be reached again, being much more veered away from **A**($|\phi_1^+\rangle$) and **a**($|\psi_1\rangle$), possibly up to even *orthogonal* to $|\psi_1\rangle$ (c.f. the state vector "**R**" in Fig. A.1). Indeed, if $|\psi_0\rangle$ and $|\psi_1\rangle$ are well reached, and if $\Psi_\perp = (\alpha_2\psi_2 + \alpha_3\psi_3 + ...)\big/\sqrt{\alpha_2^2 + \alpha_3^2 + ...}$ is normalized and orthogonal to both, i.e.

$\Psi_\perp = (\Psi - \alpha_0 \psi_0 - \alpha_1 \psi_1)/\sqrt{1 - \alpha_0^2 - \alpha_1^2}$, where $\alpha_0 = \langle \psi_0 | \Psi \rangle$, $\alpha_1 = \langle \psi_1 | \Psi \rangle$ and $\Psi$ is an arbitrary function $\Psi = (\alpha_0 \psi_0 + \alpha_1 \psi_1 + \alpha_2 \psi_2 + \alpha_3 \psi_3 + ...)/\sqrt{\alpha_0^2 + \alpha_1^2 + \alpha_2^2 + \alpha_3^2 + ...}$, then the following function $\Phi$ (by construction orthogonal to $\psi_1$)

$$\Phi = \sqrt{\frac{E[\Psi_\perp] - E_1}{E[\Psi_\perp] - E_0}} \psi_0 + 0\psi_1 - \sqrt{\frac{E_1 - E_0}{E[\Psi_\perp] - E_0}} \Psi_\perp$$

$$\Phi = \frac{\sqrt{E[\Psi] + \alpha_0^2(E_1 - E_0) - E_1} \psi_0 - \sqrt{E_1 - E_0}(\Psi - \alpha_0 \psi_0 - \alpha_1 \psi_1)}{\sqrt{E[\Psi] - \alpha_1^2(E_1 - E_0) - E_0}}$$

will have energy $E[\Phi] = E_1$, that is **$\Phi$ will have the energy of $\psi_1$ while being orthogonal to $\psi_1$**. (Since $E_1$ is known, we could **engineer** $\Psi$, and we could manipulate a chemical reaction occurring at energy $E_1$ **without using** the state $\psi_1$, i.e. by using more remote electrons comprising $\Psi_\perp = (\alpha_2 \psi_2 + \alpha_3 \psi_3 + ...)/\sqrt{\alpha_2^2 + \alpha_3^2 + ...}$). For example, putting as $\Psi$, above, the function $\Phi$100H of TABLE 1, with $E[\Psi] = -2.07215$ a.u., $\alpha_0 = 1.97\ 10^{-5}$, $\alpha_1 = 0.875$, and $\langle \psi_1 | \psi_0 \rangle = 1.72\ 10^{-4} \approx 0$, the function

$$\Phi = 0.541529\ \psi_0 + 1.51948\ \psi_1 - 1.73655\ \Psi$$

is normalized, has $E[\Phi] \approx -2.14584$ a.u. $= E_1$ and is orthogonal to $\psi_1$, having $\langle \psi_1 | \Phi \rangle \approx 0$.

The functional $F_1$ (minimization principle for excited states) approaches freely **a**($|\psi_1\rangle$), without any orthogonality constraint to crude lower lying approximants **G**($|\phi_0\rangle$).

**Acknowledgement**
This work was sponsored by: Polynano-Kripis 447963 / GSRT, Greece; partially presented in ICCMSE2015, 20-23 March 2015, Athens, Greece